\documentclass[twocolumn]{aastex631}
\usepackage{CJKutf8}
\graphicspath{{./},{./figures/}}
\usepackage{amsmath}
\usepackage{mathptmx,txfonts}
\usepackage{booktabs}
\usepackage{xspace}
\usepackage[flushleft]{threeparttable}
\usepackage{bm}
\usepackage{tablefootnote}
\usepackage[capitalize]{cleveref}

\newcommand{\pbjam}{{\texttt{PBJam}}\xspace}
\newcommand{\reggae}{{\texttt{reggae}}\xspace}

\begin{document}

\correspondingauthor{Christopher Lindsay}
\email{christopher.lindsay@yale.edu}

\title{Precise Asteroseismic Ages for the Helmi Streams}

\author{Christopher J. Lindsay\,\href{https://orcid.org/0000-0001-8722-1436}{\includegraphics[scale=0.04]{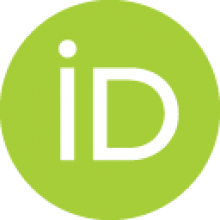}}}
\affiliation{Department of Astronomy, Yale University, PO Box 208101, New Haven, CT 06520-8101, USA}

\author{Marc Hon\,\href{https://orcid.org/0000-0003-2400-6960}{\includegraphics[scale=0.04]{orcid-ID.png}}}
\affiliation{Kavli Institute for Astrophysics and Space Research, Massachusetts Institute of Technology, Cambridge, MA 02139, USA}

\author{J. M. Joel Ong \begin{CJK*}{UTF8}{gbsn}(王加冕)\end{CJK*}\,\href{https://orcid.org/0000-0001-7664-648X}{\includegraphics[scale=0.04]{orcid-ID.png}}}
\affiliation{Institute for Astronomy, University of Hawai`i, 2680 Woodlawn Drive, Honolulu, HI 96822, USA}
\affiliation{Hubble Fellow}

\author{Rafael A. Garc\'\i a\,\href{https://orcid.org/0000-0002-8854-3776}{\includegraphics[scale=0.04]{orcid-ID.png}}} 
\affiliation{Universit\'e Paris-Saclay, Universit\'e Paris Cit\'e, CEA, CNRS, AIM, 91191, Gif-sur-Yvette, France}

\author{Dinil B. Palakkatharappil\,\href{https://orcid.org/0000-0002-6812-4443}{\includegraphics[scale=0.04]{orcid-ID.png}}} 
\affiliation{Universit\'e Paris-Saclay, Universit\'e Paris Cit\'e, CEA, CNRS, AIM, 91191, Gif-sur-Yvette, France}

\author{Jie Yu\,\href{https://orcid.org/0000-0002-0007-6211}{\includegraphics[scale=0.04]{orcid-ID.png}}} 
\affiliation{School of Computing, Australian National University, Acton, ACT 2601, Australia}
\affiliation{Research School of Astronomy \& Astrophysics, Australian National University, Cotter Road, Weston, ACT 2611, Australia}

\author{Tanda Li  \begin{CJK*}{UTF8}{gbsn}(李坦达)\end{CJK*}\,\href{https://orcid.org/0000-0001-6396-2563}{\includegraphics[scale=0.04]{orcid-ID.png}}}
\affiliation{Department of Astronomy, Beijing Normal University, Haidian District, Beijing 100875, People's Republic of China}

\author{Tomás Ruiz-Lara\,\href{https://orcid.org/0000-0001-6984-4795}{\includegraphics[scale=0.04]{orcid-ID.png}}}
\affiliation{Departamento de Física Téorica y del Cosmos, Universidad de Granada, 18071 Granada, Spain}
\affiliation{Instituto Carlos I de Física Téorica y Computacional, Facultad de Ciencias, 18071 Granada, Spain}

\author{Amina Helmi\,\href{https://orcid.org/0000-0003-3937-7641}{\includegraphics[scale=0.04]{orcid-ID.png}}}
\affiliation{Kapteyn Astronomical Institute, University of Groningen, Landleven 12, 9747 AD Groningen, The Netherlands}


\shortauthors{Lindsay, Hon et al.}
\shorttitle{Helmi Streams Erythrogigantoacoustics}



\begin{abstract}
The Helmi streams are remnants of a dwarf galaxy that was accreted by the Milky Way and whose stars now form a distinct kinematic and chemical substructure in the Galactic halo. Precisely age-dating these typically faint stars of extragalactic origin has been notoriously difficult due to the limitations of using only spectroscopic data, interferometry, or coarse asteroseismic measurements. Using observations from NASA's Transiting Exoplanet Survey Satellite, we report the detailed asteroseismic modeling of two of the brightest red giants within the Helmi streams, HD 175305 and HD 128279. By modeling the individual oscillation mode frequencies and the spectroscopic properties of both stars, we determine their fundamental properties including mass, radius, and age ($\tau$). We report $\tau = 11.16 \pm 0.91$ Gyr for HD 175305 and $\tau = 12.52 \pm 1.05$ Gyr for HD 128279, consistent with previously inferred star-formation histories for the Helmi streams and the differential chemical abundances between the two stars. With precise ages for individual stream members, our results reinforce the hypothesis that the Helmi streams' progenitor must have existed at least 12 Gyr ago. Our results also highlight that the ages of metal-poor, $\alpha$-enhanced red giants can be severely underestimated when inferred using global asteroseismic parameters instead of individual mode frequencies.  \end{abstract}



\section{Introduction} \label{sec:intro}

The $\Lambda$CDM cosmological model predicts that our galaxy has grown hierarchically through mergers with smaller satellite galaxies \citep{Springel2005}. Observational evidence supporting this scenario comes from the dynamics and compositions of the Milky Way's halo-cluster system, as halo star abundance patterns align with the idea that galaxies build their stellar populations by accreting gas and already-formed stars from lower-mass galaxies \citep{Searle_Zinn1978, Bland-Hawthorn2002, Richter2017}. The goal of Galactic archaeology is to build a time-resolved picture of the Milky Way's formation, including its early merger events. To that end, the \textit{Gaia} mission \citep{Gaia_mission, Gaia_DR2, Gaia_DR3} has successfully enabled the detection of kinematically distinct stellar-population structures within the Milky Way as evidence for both minor and major mergers \citep[e.g.,][]{Grillmair2016, Helmi2018, Helmi2020}. 

Even though stars accreted in the distant past have since phase-mixed with the Milky Way's stars, dispersing in position and velocity thus blending into the galaxy's overall stellar population, their structure remains preserved in the space of integrals of motion \citep{Helmi2000, Helmi2020, Dodd2023}. By combining position and velocity measurements from \textit{Gaia} with stellar abundance data, it is possible to determine a star's progenitor. Progenitor identification within the \textit{Gaia} 6D sample has already been carried out \citep[e.g.,][]{Koppelman2018, Price-Whelan2018}, revealing numerous structures in the stellar halo, including the Gaia-Enceladus/Sausage, which played a major role in forming the Milky Way's inner halo \citep[][and references therein]{Helmi2018}. Both the Gaia-Enceladus Sausage \citep{Koppelman2018} and Sagittarius streams \citep{Ibata1994} form large contributions to the Milky Way’s inner halo and also influence the star formation history of the Milky Way \citep{Ruiz-Lara2020}. The inclusion of individual elemental abundance measurements reveals that individual streams may have distinct chemical properties indicative of their progenitors' star formation history \citep[e.g.,][]{Roederer2010, Naidu2020, Ruiz-Lara2022_IOM, Matsuno2022, Dodd2023, Horta2023}.

While the identification of stellar streams has been greatly invigorated in recent years \citep{Bonaca2025}, the exact history of their merging into the Galaxy remains highly uncertain. Traditional approaches for determining the ages of stellar streams applied isochrone fitting, which may be particularly uncertain for red giants (e.g., \citealt{Das_2020}). The advent of asteroseismology as a tool for Galactic archaeology has provided new prospects for precisely age-dating stellar populations around the Milky Way, including members of stellar streams. 

The long temporal baselines from recent space based missions such as \textit{Kepler} \citep{Kepler_inst} and TESS \citep{TESS_inst} permit the frequency resolution required for determining the frequencies of individual oscillation modes in the power spectra of many stars \citep{Garcia_review_2019}. Stars with convective envelopes, such as our Sun or evolved low-mass stars, oscillate in many frequencies simultaneously \citep{Goldreich1977a,Goldreich1977b}. Measurements of the oscillation mode properties allow for detailed studies of stellar interiors, as well as precise determinations of the global stellar properties like mass, radius, and age. This is because the properties a star's convectively excited oscillation modes are governed by the star's interior structure, and therefore change with stellar properties and age.

The oscillation power of evolved low-mass stars is concentrated in a Gaussian-shaped envelope around the frequency of maximum oscillation power ($\nu_{\text{max}}$), while the individual oscillation modes appear as peaks of power superimposed on this envelope. For oscillation modes where the restoring force is pressure, modes of the same angular degree $\ell$ are regularly spaced in frequency by a typical separation called the large frequency separation \citep[$\Delta \nu$,][]{Kjeldsen1995, mybook}.

Thus far, nearly all asteroseismic studies involving Galactic archaeology, rely on the two `global' asteroseismic measurements,  $\Delta \nu$ and $\nu_{\text{max}}$, determined from the oscillation power spectra of low-mass red giants \citep{Miglio2013, Huber2019_review, Miglio2021}. Because $\Delta \nu$ scales approximately with the square root of the stellar density, while $\nu_{\text{max}}$ scales with the star’s surface gravity and temperature, these global asteroseismic measurements provide constraints on red giant radii, masses, and subsequently ages when combined with temperature measurements. Previous works have determined $\Delta \nu$ and $\nu_{\text{max}}$ measurements for remnant stars born in external galaxies \citep{Borre2022} as well as many other low metallicity halo and thick disk stars observed by TESS \citep{Grunblatt2021, Marasco2025}. 

Global asteroseismology using scaling relations is applicable to thousands of metal-poor stars, however, there has been mounting evidence that the use of global asteroseismic measurements for metal-poor stars significantly overestimates masses compared with stellar masses determined using a stars individual oscillation mode frequencies \citep[e.g.][]{Huber2024, Larsen2025}. The structure of evolved stars permits non-radial ($\ell > 0$) oscillation modes of mixed character, which are a result of coupling between gravity modes, which propagate in evolved stars cores, and pressure modes, which propagate in stellar envelopes \citep{Osaki1975, Aizenman1977, Shibahashi1979, Takata_2016a}. Since they can probe both the deep core and outer layers of stars, mixed mode oscillation frequencies are valuable inputs for modeling evolved stars and precisely determining stellar ages. This discrepancy is not observed when individual oscillation modes are fitted to stellar models, which remains the gold standard for asteroseismic analyses. Detailed asteroseismic modeling has not yet been performed for any star that is firmly established to be a member of a stellar stream. The detailed asteroseismic modeling of $\nu\,$Indi \citep{Chaplin2020} provided an exemplary study for age-dating the Gaia-Enceladus merger, though the star is thought to have been born in-situ.  

In this work, we study two firmly established members of the Helmi streams: red giants HD 175305 and HD 128279. The Helmi streams are the remnants of a disrupted dwarf galaxy with an estimated stellar mass of about $10^8 \text{M}_{\odot}$, believed to have merged with the Milky Way approximately 5 to 8 Gyr ago \citep{Kepley2007, Koppelman2018, Naidu2022, Woudenberg2024}. Having many stream members within several kiloparsecs to the Sun, the Helmi streams were among the earliest kinematic substructures associated with an accretion event \citep{Helmi1999}. It is this proximity which provides us with the novel opportunity to perform the detailed asteroseismic modeling of the two red giants, which are thought to be of extragalactic origin.

We report the spectroscopic and asteroseismic data for the two target stars in \autoref{sec:target_stars} and describe our asteroseismic modeling methods in \autoref{sec:optimization}. The modeling results are detailed in \autoref{sec:results} and implications of these results are discussed in \autoref{sec:discussion}.

\begin{table*}[ht]
    \centering
    \caption{The stellar parameters of the target stars analyzed in this study, HD 175305 and HD 128279.}\label{table:stellar_parameters}
    \begin{threeparttable}
    \begin{tabular}{ccccc}
        \toprule
        & \multicolumn{2}{c}{\textbf{HD 175305}} & \multicolumn{2}{c}{\textbf{\textbf{HD 128279}}} \\
        \midrule
        & Value & Source & Value & Source \\
        \midrule
        TIC ID & 335965870 & TESS Input Catalog\textsuperscript{*} &13727 & TESS Input Catalog\textsuperscript{*}\\       
        \textit{Gaia} G Magnitude &$6.96$ & \textit{Gaia} DR3\textsuperscript{\textdaggerdbl} &$7.82$ & \textit{Gaia} DR3\textsuperscript{\textdaggerdbl}\\       
        Effective Temperature (K) & $5036\pm200$ & \citet{Ishigaki_2012} &$5328\pm200$ & \citet{Ishigaki_2012}\\
        Luminosity ($\text{L}_{\odot}$) & $33.1\pm3.0$ & This work & $11.1 \pm 1.7$ & This work \\
        Parallax (mas) & $6.41\pm 0.01$ &\textit{Gaia} DR3\textsuperscript{\textdaggerdbl} &$7.63\pm 0.03$ & \textit{Gaia} DR3\textsuperscript{\textdaggerdbl}\\
        $[$Fe/H$]$ (dex) & $-1.35\pm0.15$ & \citet{Ishigaki_2012} & $-2.17\pm0.12$&\citet{Ishigaki_2012}\\
        $[\alpha$/Fe$]$ (dex) & $0.23\pm0.05$ & \citet{Ishigaki_2012}  & $0.30\pm0.05$&\citet{Ishigaki_2012}\\
        $\Delta\nu$ ($\mu$Hz) & $5.89\pm0.01$ & This work &$15.87\pm0.05$&This work\\
        $\nu_{\mathrm{max}}$ ($\mu$Hz) & $52.17\pm0.37$ & This work  & $189.10\pm0.52$&This work\\
        \bottomrule
    \end{tabular}
\begin{tablenotes}
\item[*] \citet{Stassun_2019}
\item[\textdaggerdbl] \citet{Gaia_DR3}
\end{tablenotes}
    \end{threeparttable}
\end{table*} 


\section{Target Stars} \label{sec:target_stars}

The spectroscopic and global asteroseismic properties of the target stars, HD 175305 and HD 128279 are listed in \autoref{table:stellar_parameters} along with the data sources. HD 175305 is a metal poor ([Fe/H] = -1.35) and moderately alpha-enhanced ([$\alpha$/Fe] = 0.23) red giant star in the Transiting Exoplanet Survey Satellite's \citep[TESS][]{TESS_inst} Northern Continuous Viewing Zone.  It is also a Gaia-benchmark star, with its elemental abundances and fundamental properties studied extensively using spectroscopy (e.g., \citealt{Ishigaki_2012, APOGEE_DR17}) and interferometry \citep{Karovicova2020, Soubiran_2024}. HD 128279 is also a red giant star, but is located closer to the ecliptic. Compared to HD 175305, HD 128279 is more metal poor ([Fe/H] = -2.17), more alpha enhanced ([$\alpha$/Fe] = 0.30), and less luminous. The luminosity values we report for both HD 175305 and HD 128279 were calculated using the SED EXplorer spectral energy distribution fitting pipeline \citep[SEDEX][]{Yu_2023}.



\subsection{Kinematics}
\begin{figure}
    \centering
    \includegraphics[width=0.47\textwidth]{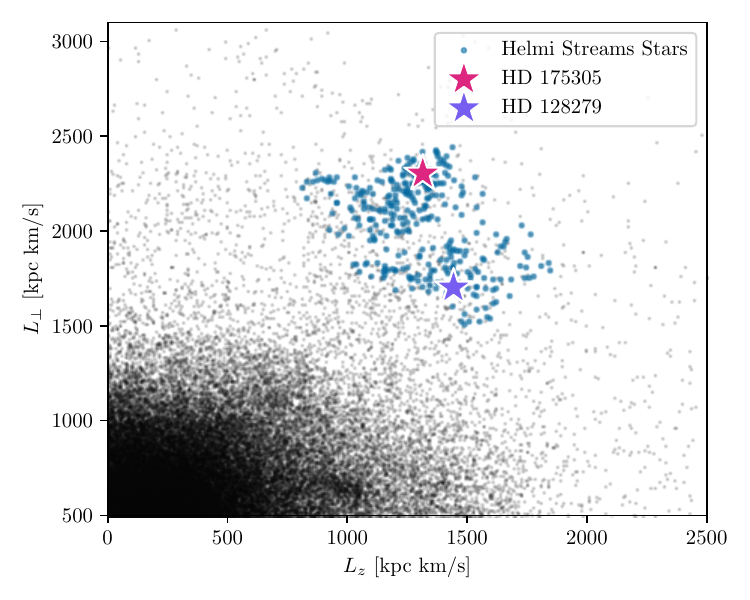}
    \caption{Black points show the kinematically selected Halo star sample from \citet{Dodd2023} in $L_{\mathrm{\perp}}$ versus the $L_{\mathrm{z}}$ space. The groupings of stars associated with the Helmi streams and our target stars are marked with blue points and star symbols respectively. HD 175305 (dark red star) and HD 128279 (purple star) are members of the high $L_{\perp}$ and low $L_{\perp}$ clumps described in \citet{Woudenberg2024} respectively. }
    \label{fig:Lz_Lperp}
\end{figure}

HD 175305 and HD 128279 were first identified as members of the Helmi streams using kinematic measurements from the \textit{Hipparcos} satellite \citep{Helmi1999}, forming part of a distinct clustering in angular momentum ($L_z-L_{\perp}$) space. Subsequent work analyzing the kinematics of metal poor stars \citep{Beers_2000, Chiba_2000} also confirmed HD 175305 and HD 128279's membership in the kinematic substructure described in \citet{Helmi1999}. These two stars have maintained their stream membership in the presence of more comprehensive 5-D and 6-D kinematic data from the \textit{Gaia} mission \citep{Koppelman2019, Lovdal2022, Ruiz-Lara2022_IOM, Dodd2023}. \autoref{fig:Lz_Lperp} shows the sample of kinematically selected Halo stars from \citet{Dodd2023} in black points and the Helmi Streams stars showed with blue points in $L_{\mathrm{\perp}}$ versus the $L_{\mathrm{z}}$ space. \citet{Woudenberg2024} pointed out that the collection of Helmi streams stars identified in \citet{Dodd2023} shown in can be split into two separate clumps in ($L_z, L_{\perp}$) space. \autoref{fig:Lz_Lperp} shows that HD 175305 (marked with a dark red star symbol) is a member of the high-$L_{\perp}$ grouping while HD 128279 (marked with a purple star symbol) is a member of the low-$L_{\perp}$ clump. HD 175305 and the other high-$L_{\perp}$ stars have larger $z$-velocities (velocity out of the plane of the galaxy). Within the high-$L_{\perp}$ clump within the Helmi streams, previous work has identified a part of the Helmi streams that is significantly less phased mixed, dubbed the subclump in \citet{Woudenberg2024}. We note that HD 175305 and HD 128279 are not identified as members of the subclump.

\subsection{Asteroseismic Mode Determination}

The multiple sectors of TESS light curves for both our target stars were combined and corrected for systematics. The light curve for HD 175305 was generated using the MIT Quick-Look Pipeline \citep[QLP;][]{Quicklook1, Quicklook2, Kunimoto2021, Kunimoto2022}, which provides sufficient precision for observing giant star oscillations \citep{Hon2021QuickLook}. We used QLP light curves from TESS sectors 14-26, 41, 47-51, 53-60, and 73-74 that were processed further using the PYTADACS (Python for Tess Astroseismic and DynAmiC Studies) pipeline (Garc\'\i a et al. in prep). The pipeline stitches together contiguous light curve segments spaced further than 3 consecutive sectors apart and performs two iterations of $\sigma$-clipping. The first iteration removes all flux measurements deviating further than 10$\,\sigma$ of the light curve's mean flux, while the second iteration removes (from the original light curve), flux measurements deviating further than $4\,\sigma$ from the data smoothed with a median filter of 1-day width. Next, the light curve is binned into a regular grid with a 30-minute cadence using the nearest neighbor resampling algorithm with the slotting principle \citep{Broersen_2009}, as described in \citealt{Garcia_2014, Garcia2024}. Finally, the light curve is high-pass filtered using a 5-day triangular filter. The effect of the gap removal prescription applied in the light curve preparation we used was studied in \citet{Cuesta2024}.

We used the light curve generated by the TESS Science Processing Operations Center pipeline \citep[SPOC;][]{Twicken2016, Jenkins2017} for HD 128279. Though HD 175305 does not have SPOC light curves, we used SPOC data for HD 128279 because they generally preserve flux variability better for bright red giants \citep{Hon2022HDTESS}. 

\begin{figure*}[htbp]
    \centering
    \includegraphics[width=0.45\textwidth]{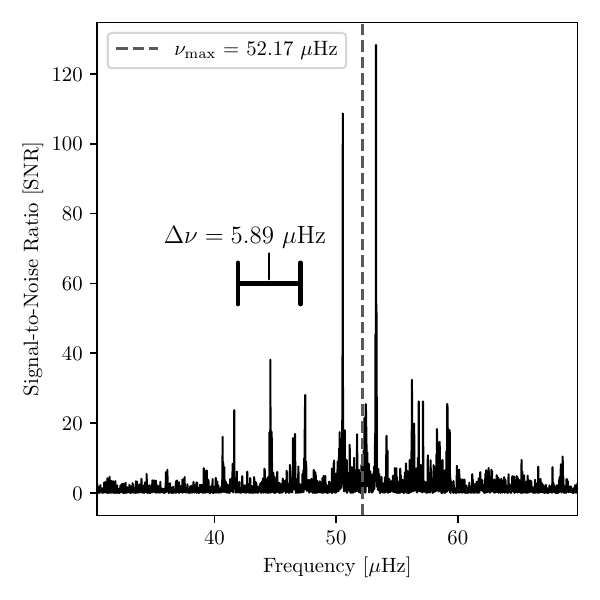}
    \hspace{0.01\textwidth}
    \includegraphics[width=0.45\textwidth]{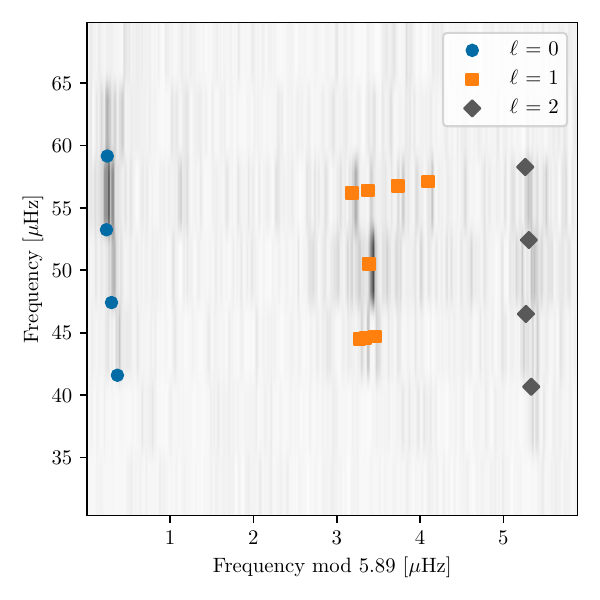}
    \caption{HD 175305 background-subtracted power spectral density (left) and power échelle plot (right). The markers in the power échelle plot show the observed mode frequencies determined using TACO \citep{taco}.  }
    \label{fig:HD175305_peakbagging}
\end{figure*}

\begin{figure*}[htbp]
    \centering
    \includegraphics[width=0.45\textwidth]{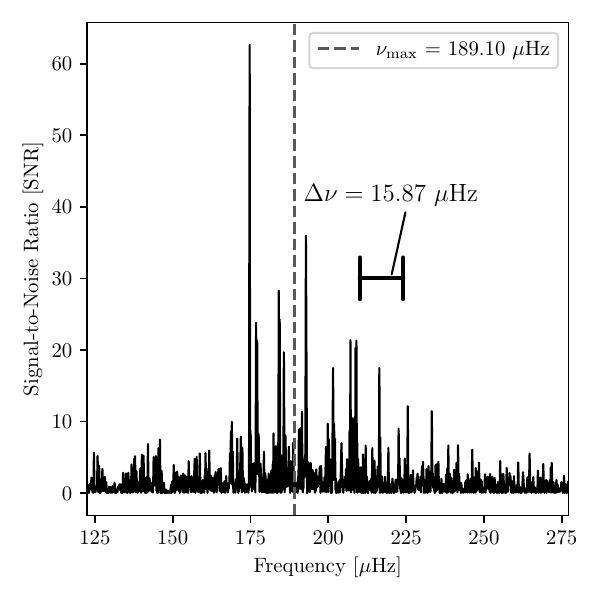}
    \hspace{0.01\textwidth}
    \includegraphics[width=0.45\textwidth]{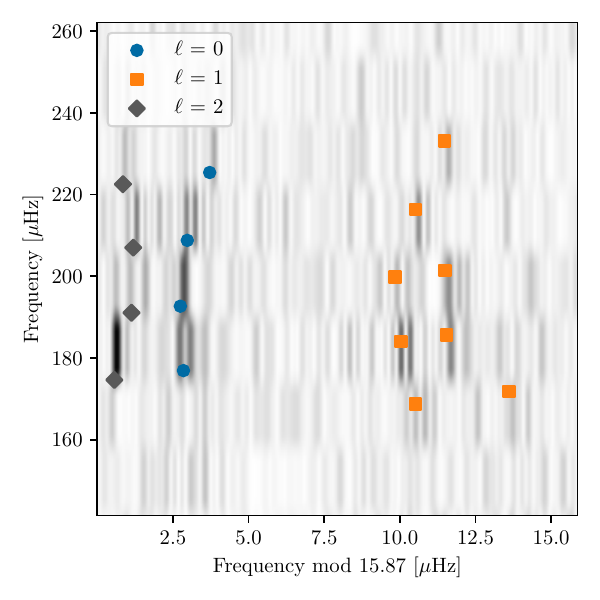}
    \caption{HD 128279 Background-subtracted power spectral density (left) and power échelle plot (right). The markers in the power échelle plot show the observed mode frequencies determined using TACO \citep{taco}. }
    \label{fig:HD128279_peakbagging}
\end{figure*}

The light curves of both target stars were analyzed for stellar oscillations using the TACO (Tools for Automated Characterization of Oscillations) code \citep{taco}. The global asteroseismic parameters, $\Delta \nu$ and $\nu_{\text{max}}$, we report for HD 175305 and HD 128279 in \autoref{table:stellar_parameters} were calculated using TACO by first calculating a power density spectrum (PSD) from the light curve, dividing out the background component from convective granulation and white noise background, and finding the frequency of maximum oscillation power ($\nu_{\text{max}}$). We show the background-subtracted power spectral density for HD 175305 and HD 128279 in the left panels of \autoref{fig:HD175305_peakbagging} and \autoref{fig:HD128279_peakbagging} respectively. 

Peaks are then identified from the PSD by applying a Mexican-hat wavelet-transform based algorithm iteratively to find resolved peaks, which are then fitted with Lorentzian functions. The $\ell = 0$ and $\ell = 2$ modes are identified by using the universal pattern for solar-like p-mode oscillations of the same angular degree, $\nu_{n_p, \ell, m} \sim \Delta \nu \big(n_p + \frac{\ell}{2} + \epsilon_p\big)$, where $\Delta \nu$ is the large frequency separation, $\epsilon_p$ is a phase term, and $n_p$, $\ell$, and $m$ are the p-mode radial order, angular degree, and azimuthal order respectively \citep{Tassoul1980}. $\Delta \nu$ is identified by TACO using the frequency spacing between consecutive radial ($\ell = 0$) p-modes. For the peaks not identified as $\ell = 0$ or $\ell = 2$ modes, TACO identifies them as dipole modes ($\ell = 1$). The mode frequencies, mode frequency errors, and angular degrees for HD 175305 and HD 128279 are reported in \autoref{table:HD_175305_freqs} and \autoref{table:HD_128279_freqs} respectively. The mode frequencies errors were determined by TACO, but we adopt the frequency resolution of the power spectrum (1 divided by the length of the light curve) as a lower bound for the frequency error. The lower limits for the mode frequency errors are 0.0152 $\mu$Hz for HD 175305 and 0.234 $\mu$Hz for HD 128279. 

Note that the mode frequencies and mode frequency errors were corrected for stellar line-of-sight Doppler velocity shifts following \citet{Davies2014} using radial velocities from GAIA DR3 \citep[-184.08$\pm$0.12 km/s for HD 175305 and -75.68$\pm$0.14 km/s for HD 128279, ][]{Gaia_DR3}. This is done by multiplying the observed frequency ($\nu_0$) and frequency uncertainty ($\sigma_{\nu_0}$) by a factor $\sqrt{\frac{1 + \beta}{1 - \beta}} \simeq (1 + \beta)$ to obtain the frequency emitted by the source ($\nu_s$) and uncertainty in Doppler shifted pulsations ($\sigma_{\nu_s}$). For HD 175305, $\frac{\nu_s}{\nu_0} = \frac{\sigma_{\nu_s}}{\sigma_{\nu_0}} \simeq 0.9994$ while for HD 128279, $\frac{\nu_s}{\nu_0} = \frac{\sigma_{\nu_s}}{\sigma_{\nu_0}} \simeq 0.9997$. 

The mode frequencies are visualized using frequency échelle diagrams in the right panels of \autoref{fig:HD175305_peakbagging} and \autoref{fig:HD128279_peakbagging} for HD 175305 and HD 128279 respectively. The background power échelle diagrams are constructed, following \citet{Grec1983}, by splitting a star's oscillation power spectrum into chunks of width $\Delta \nu$, then stacking these chunks in order of increasing frequency. Although the frequency échelle diagrams of both HD 175305 and HD 128279 exhibit the presence of dipolar mixed modes (more than one $\ell = 1$ mode per radial order) the lower $\Delta \nu$ value and denser dipole mode spectrum in HD 175305's échelle diagram indicates that HD 175305 is more evolved compared with HD 128279. The presence of dipolar mixed modes indicates that both stars are evolved. In the rest of our analysis, we assume that both HD 17305 and HD 128279 are first ascent red giant branch stars, with hydrogen nuclear burning occurring in a shell around an inert helium core. We further discuss the evolutionary state of HD 175305 in \autoref{subsec:175305_State}.

\begin{table}[ht!]
\caption{Observed Mode Frequencies for HD 175305 determined using the Tools for Automated Characterization of Oscillations (TACO) Code \citep{taco}. }
\label{table:HD_175305_freqs}
\centering
\begin{tabular}{ccc}
\toprule
Angular Degree ($\ell$) & Frequency ($\mu$Hz) & Error ($\mu$Hz) \\
\midrule
2                                           & 40.67                                   & 0.02                                \\
0                                           & 41.60                                   & 0.02                                \\
1                                           & 44.50                                   & 0.02                                \\
1                                           & 44.57                                   & 0.02                               \\
1                                           & 44.69                                   & 0.02                                \\
2                                           & 46.50                                   & 0.04                                \\
0                                           & 47.42                                   & 0.02                                \\
1                                           & 50.51                                   & 0.02                                \\
2                                           & 52.43                                   & 0.04                                \\
0                                           & 53.25                                   & 0.02                                \\
1                                           & 56.19                                   & 0.02                                \\
1                                           & 56.38                                   & 0.02                                \\
1                                           & 56.74                                   & 0.03                                \\
1                                           & 57.11                                   & 0.02                                \\
2                                           & 58.28                                   & 0.04                                \\
0                                           & 59.15                                   & 0.03                               \\
\bottomrule
\end{tabular}
\end{table}

\begin{table}[ht!]
\caption{Observed Mode Frequencies for HD 128279 determined using TACO. }
\label{table:HD_128279_freqs}
\centering
\begin{tabular}{ccc}
\toprule
Angular Degree ($\ell$) & Frequency ($\mu$Hz) & Error ($\mu$Hz) \\
\midrule
1                       & 168.72              & 0.23                     \\
1                       & 171.81              & 0.23                     \\
2                       & 174.63              & 0.23                     \\
0                       & 176.92              & 0.23                     \\
1                       & 184.10              & 0.23                     \\
1                       & 185.62              & 0.23                     \\
2                       & 191.07              & 0.23                     \\
0                       & 192.69              & 0.23                     \\
1                       & 199.78              & 0.23                     \\
1                       & 201.42              & 0.23                     \\
2                       & 207.00              & 0.23                     \\
0                       & 208.78              & 0.23                     \\
1                       & 216.32              & 0.23                     \\
2                       & 222.53              & 0.23                     \\
0                       & 225.39              & 0.23                     \\
1                       & 233.15              & 0.23                    \\
\bottomrule
\end{tabular}
\end{table}

\section{Asteroseismic Optimization} \label{sec:optimization}

In our modeling of HD 175305 and HD 128279, we used the effective temperature, [Fe/H], and [$\alpha$/Fe] measurements from \citet{Ishigaki_2012} listed in \autoref{table:stellar_parameters} along with luminosities and the individual oscillation mode frequencies tabulated in \autoref{table:HD_175305_freqs} and \autoref{table:HD_128279_freqs}. The asteroseismic optimization pipeline developed for this project employs the stellar evolution code MESA \citep[Modules for Experiments in Stellar Astrophysics][]{Paxton2011, Paxton2013, Paxton2015, Paxton2018, Paxton2019, Jermyn2023} and the differential evolution algorithm implemented in \texttt{yabox} \citep{Yabox} in order to perform on-the-fly stellar modeling and find stellar model parameters that minimize a cost function taking into account the models' match to the spectroscopic and asteroseismic observables. The parameters we vary are initial mass ($M_0$, initial helium mass fraction $Y_0$, initial metallicity ([Fe/H$_0$]), and convective mixing length ($\alpha_{\text{mlt}}$). Using \texttt{yabox} allows for derivative-free minimization of a cost function, which is evaluated for different combinations of the initial mass, metal abundance, helium abundance and convective mixing length in the steps detailed in \autoref{appendix1}. The model cost function is similar to that defined in \citet{Lindsay2024} and incorporates the fit to the spectroscopic observables as well as the fit between the model mode frequencies corrected for near surface effects and the observed frequencies \citep{bg14}. 



There is not a specific cost value limit for which the algorithm stops, instead the differential evolution process continues for a set number of iterations (40 iterations, 840 cost function evaluations) and the minimum output of the cost function through all the evaluations is taken as the best-fit track, with the model corresponding to the minimum $\chi^2_{\text{total}}$ value along the track being the best-fit model. The use of the \texttt{yabox} differential evolution algorithm ensures that the region of parameter space around the highest likelihood initial parameters is more densely sampled, compared with uniform sampling as is implemented in most grid based modeling studies. 

The age of the universe ($\tau_{\textrm{universe}} \simeq 13.8$ Gyr) was not explicitly used as a cutoff in the MESA evolution, however an age prior is still enforced through our adopted lower bound in stellar mass of $M \geq 0.70 \text{M}_{\odot}$. Many models with $M \lesssim 0.77 \text{M}_{\odot}$ have $\tau > \tau_{\textrm{universe}}$ on the red giant branch for the metallicities, luminosities, and temperatures under consideration. Therefore, there are some model tracks calculated along the optimization where the best fit model has an age greater than $\tau_{\textrm{universe}}$. However, we find that the best fit models for both HD 128279 and HD 175305 are younger than $\tau_{\textrm{universe}}$.

In addition to determining the best fit model parameters, the errors of each model parameter are estimated by taking the likelihood weighted standard deviation of each parameter. This involves first taking the best fit model from each cost function iteration track (840 in total) and multiplying the model parameter value ($P$) by the model weight ($W_{\textrm{model}}$), given by the likelihood divided by the sampling density function from the optimization track, 
\begin{equation}
W_{\textrm{model}} = \frac{L_{\textrm{total}}}{\sum L_{\textrm{total}}}, \text{ with } L_{\textrm{total}} = \exp(-\frac{1}{2} \chi^2_{\text{total}})/p(P).
\end{equation}
where $\chi^2_{\text{total}}$ is defined in \autoref{appendix1} and $p(P)$ is the Kernel Density Estimation based estimate of the local sampling density at that particular parameter value.

The likelihood weighted parameter values ($P \times W_{\textrm{model}}$) are then summed to obtain the likelihood weighted mean parameter value  ($\bar{P}_{\text{Likelihood Weighted}}$) and the likelihood weighted standard deviation of each parameter is calculated as 
\begin{equation}
\sigma_{\text{parameter}} = \sqrt{\sum W_{\textrm{model}} \times (P - \bar{P}_{\text{Likelihood Weighted}})^2}.
\end{equation}

\section{Results} \label{sec:results}

The best fit model parameters and parameter uncertainties generated by our asteroseismic modeling procedure are reported in \autoref{table:Results}. The spectroscopic $\chi^2$, seismic $\chi^2$, and low-$n$ $\chi^2$ components (defined in \autoref{appendix1}) for the best fit models of HD 175305 and HD 128279 are reported in \autoref{table:chi2}.

\begin{table*}[ht!]
\caption{Asteroseismic optimization modeling results for HD 128279 and HD 175305. The parameter values represent the best-fit model's mass, age, radius, initial helium abundance, mixing length, effective temperature, luminosity, and iron abundance. The parameter errors are determined by finding the likelihood weighted standard deviation of each parameter. The contributions to the best-fit models' $\chi^2$ from the spectroscopic, seismic, and low-$n$-value components are listed in \autoref{table:chi2}. }

\label{table:Results}
\centering
\begin{tabular}{lcccccccc}
\toprule
Target    & Mass [$\text{M}_{\odot}$] & Age [Gyr]      & Radius [$\text{R}_{\odot}$] & $Y_0$             & $\alpha_{\text{mlt}}$ & T$_{\text{eff}}$ [K] & Luminosity [$\text{L}_{\odot}$] & [Fe/H]           \\
\hline
HD 175305 & $0.83 \pm 0.02$    & $11.16 \pm 0.91$ & $7.40 \pm 0.07$      & $0.25 \pm 0.01$ & $1.80 \pm 0.10$       & $5127 \pm 68$                  & $34.10 \pm 1.58$         & $-1.46 \pm 0.13$ \\
HD 128279 & $0.77 \pm 0.02$    & $12.52 \pm 1.05$ & $3.80 \pm 0.03$      & $0.26 \pm 0.01$ & $1.97 \pm 0.03$       & $5458 \pm 39$                  & $11.54 \pm 0.39$         & $-2.17 \pm 0.10$\\
\hline
\end{tabular}
\end{table*}

\begin{table}[ht]
    \centering
    \caption{The best fit model component $\chi^2$ values. }\label{table:chi2}
    \begin{tabular}{ccccc}
        \toprule
        & \multicolumn{1}{c}{\textbf{HD 175305}} & \multicolumn{1}{c}{\textbf{\textbf{HD 128279}}} \\
        \midrule
        $\chi^2_{\textrm{spec}}$ & 0.85 & 0.49\\       
        $\chi^2_{\textrm{seismic}}$ & 2.76 & 2.09\\       
        $\chi^2_{\textrm{low n}}$ & 0.14 & 0.03\\
        \bottomrule
    \end{tabular}
\end{table}

\begin{figure*}[htbp]
    \centering
    \includegraphics[width=0.48\textwidth]{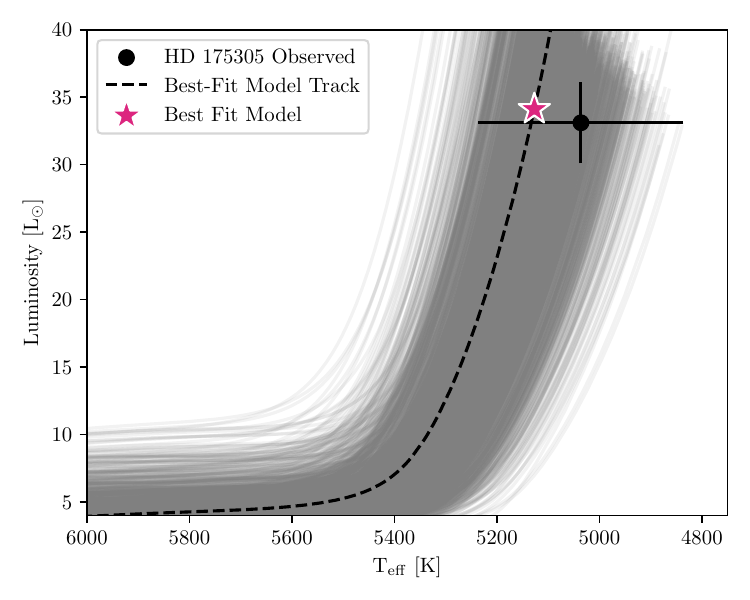}
    \hspace{0.01\textwidth}
    \includegraphics[width=0.48\textwidth]{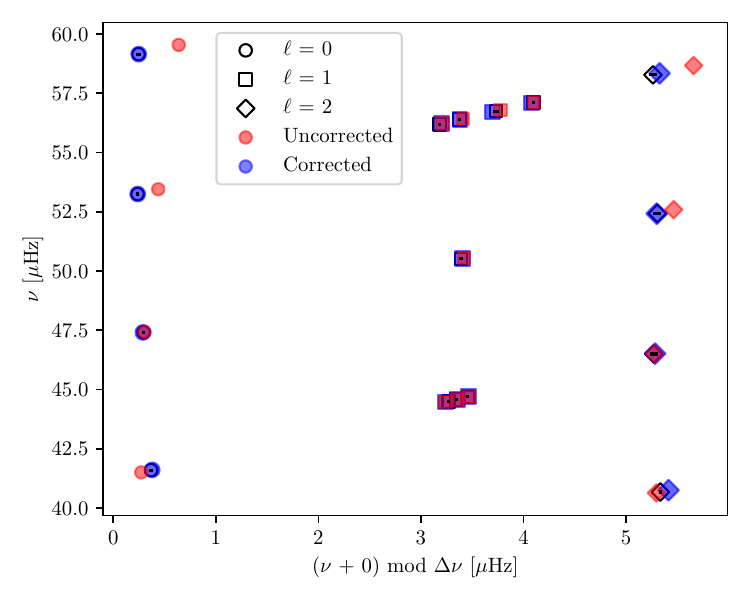}
    \caption{HD 175305 best fit model evolutionary track shown on a HR diagram (dashed line, left panel). The background evolutionary tracks on the HR diagram show the other evolutionary track calculated during the optimization process while the point with error bars display the HD 175305 observations from \citet{Ishigaki_2012}. The right panel frequency échelle diagram shows the best fit surface-corrected model oscillation modes in blue, matching the observed modes shown as open symbols. The red points show the best-fit model's non-surface-corrected modes.  }
    \label{fig:HD175305_results}
\end{figure*}

\begin{figure*}[htbp]
    \centering
    \includegraphics[width=0.48\textwidth]{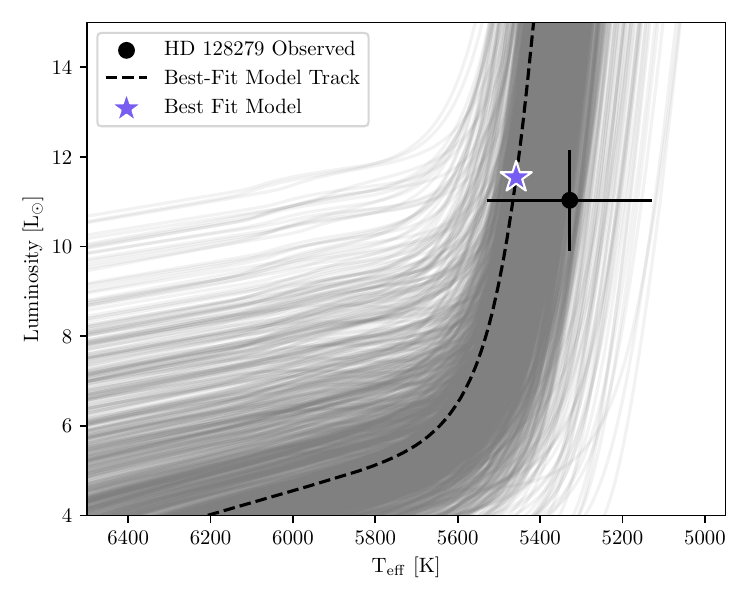}
    \hspace{0.01\textwidth}
    \includegraphics[width=0.48\textwidth]{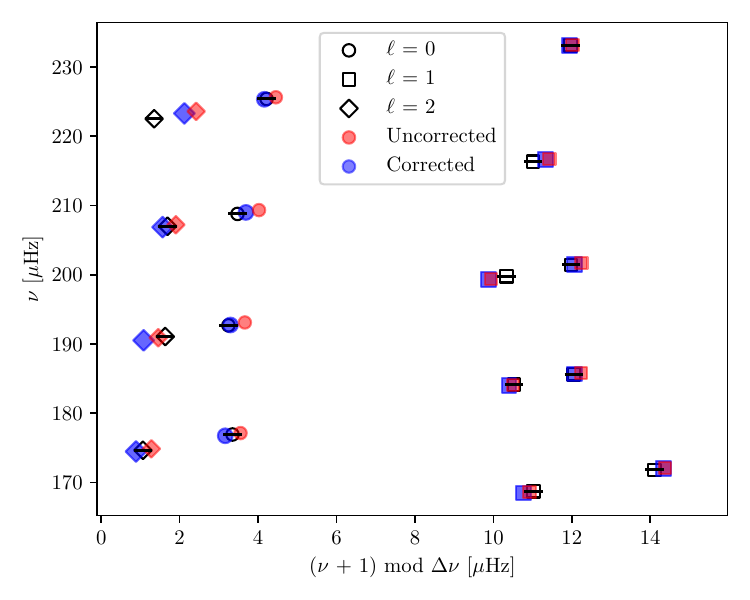}
    \caption{Same as \autoref{fig:HD175305_results} but showing the best fit model for HD 128279. }
    \label{fig:HD128279_results}
\end{figure*}

To visualize the fit between the observations from \citet{Ishigaki_2012} and the best-fit model spectroscopic parameters, the evolutionary track for the best fit models for HD 175305 and HD 128279 are shown in the left panels of \autoref{fig:HD175305_results} and \autoref{fig:HD128279_results} respectively. The background tracks in both HR-diagram panels show all the evolutionary tracks calculated over the course of the optimization process detailed in \autoref{sec:optimization}. As apparent in the left-hand panels of \autoref{fig:HD175305_results} and \autoref{fig:HD128279_results}, the best fit models' effective temperatures and luminosities agree within 1$\sigma$ with the observations. We note that for both HD 175305 and HD 128279, the best fit models' are slightly hotter, more luminous, and more metal poor compared with the observations from \citet{Ishigaki_2012}. The best fit model [Fe/H] for HD 175305 ([Fe/H]$_{\textrm{best fit}}$ = -1.46) is lower, but within 1$\,\sigma$ of the observed value ([Fe/H]$_{\textrm{observed}}$ = -1.35$\pm0.15$). The best fit model [Fe/H] for HD 128279 ([Fe/H]$_{\textrm{best fit}}$ = -2.17) agrees with the observed value ([Fe/H]$_{\textrm{observed}}$ = -2.17$\pm0.12$). 

The right panels of \autoref{fig:HD175305_results} and \autoref{fig:HD128279_results} show the best fit models' oscillation mode frequencies (filled symbols) compared with the observed mode frequencies (open symbols) determined in this work and listed in \autoref{table:HD_175305_freqs} and \autoref{table:HD_128279_freqs} for HD 175305 and HD 128279 respectively. The red symbols show the uncorrected model mode frequencies directly from the stellar oscillation code GYRE \citep{Townsend2013}, while the blue symbols show the surface-term-corrected frequencies, which are compared with the observed mode frequencies using the two term correction from \citet{bg14} (see \autoref{appendix1}).

\section{Discussion} \label{sec:discussion}
\subsection{The Star Formation History of the Helmi Streams}

The asteroseismic best-fit age for HD 175305 is $11.16 \pm 0.91$ Gyr while the asteroseimic best-fit age for HD 128279 is about 1 Gyr older, at $12.52 \pm 1.05$ Gyr. This difference in age is reflected in the older star, HD 128279, being more metal poor ([Fe/H] = -2.32) and more alpha enhanced ([$\alpha$/Fe] = 0.3) than the younger star, HD 175305, which has [Fe/H] = -1.44 and [$\alpha$/Fe] = 0.23. This is broadly consistent with known chemical enrichment trends for galaxies and the Milky Way's stellar halo, which state that older populations of stars are more metal poor and alpha enhanced. Older stars are born from material mainly enhanced from core-collapse Type II supernovae, which enrich the interstellar medium with $\alpha$-elements in higher concentrations compared with later-occurring Type Ia supernovae \citep{Tinsley1979, Roederer2010}. The Helmi streams stars are known to have a large spread in metallicities, which indicates an extended star formation history by the progenitor \citep{Koppelman2019, Naidu2020, Ruiz-Lara2022_IOM}.  

The best-fit model ages we determine for these Helmi streams members using detailed asteroseismology are consistent with the formation scenario in which a dwarf galaxy acts as the progenitor of the Helmi streams, forming a significant portion of its stars approximately 11-13 billion years ago. In particular, \citet{Koppelman2019} showed that by fitting isochrones to the population of Helmi streams stars, there exists a broad spread in metallicity ($-2.3 \leq [\text{Fe/H}] \leq -1.0$) and age (11 Gyr $\leq \text{age} \leq 13$ Gyr) that is consistent with the metallicity and age variations between HD 175305 and HD 128279. The agreement between the results of \citet{Koppelman2019}, which compared spectroscopic observations to isochrones calculated using the PARSEC stellar evolution code \citep{Bressan2012}, and our asteroseismic results, which were determined using MESA \citep{Paxton2011}, shows that broadly, the ages determined for stars in the Helmi streams are generally consistent between these different methods of analysis.  

\begin{figure}
    \centering
    \includegraphics[width=0.47\textwidth]{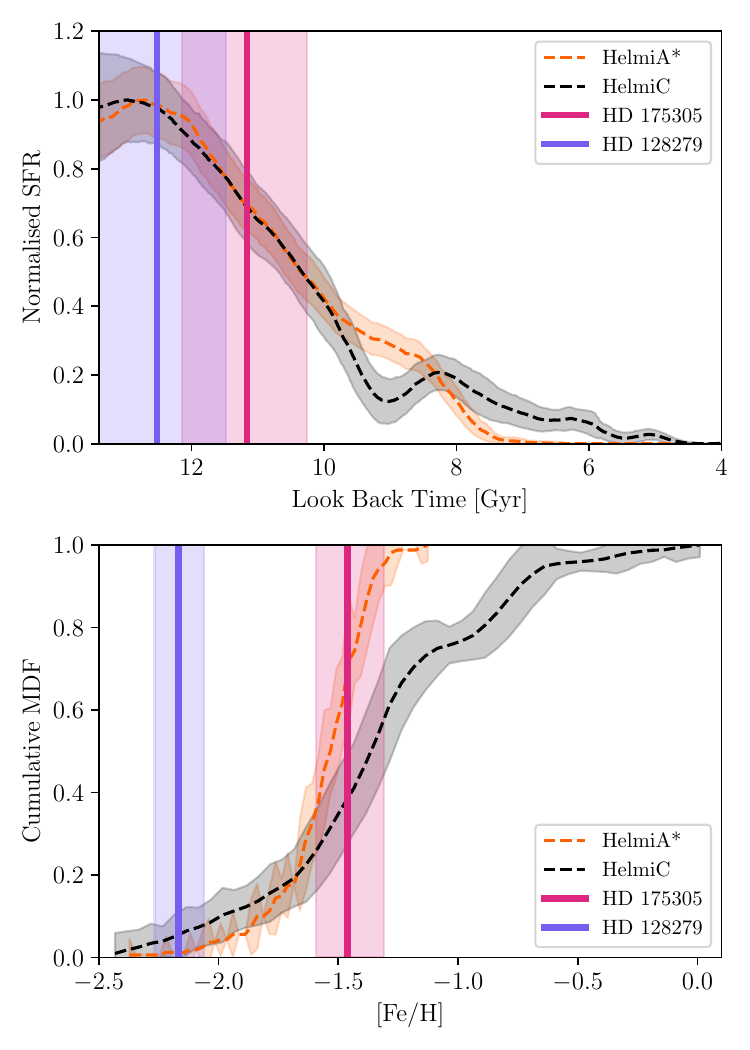}
    \caption{The top panel shows the normalized star formation rate (SFR) versus look back time results from \citet{Ruiz-Lara_helmi_history} for their 100\% purity HelmiA* subsample (red curve) and 14\% purity HelmiC subsample (black curve). 
    The bottom panel shows the cumulative metallicity distribution function (MDF) of the same Helmi streams subsamples. The thick dark red and purple vertical lines show the best fit HD 175305 and HD 128279 asteroseismic ages and [Fe/H] values in the top and bottom panels respectively, while the faint spans show the associated errors (see \autoref{table:Results}). }
    \label{fig:StarFormationHistoryFigRuizLara}
\end{figure}

\begin{figure}
    \centering
    \includegraphics[width=0.47\textwidth]{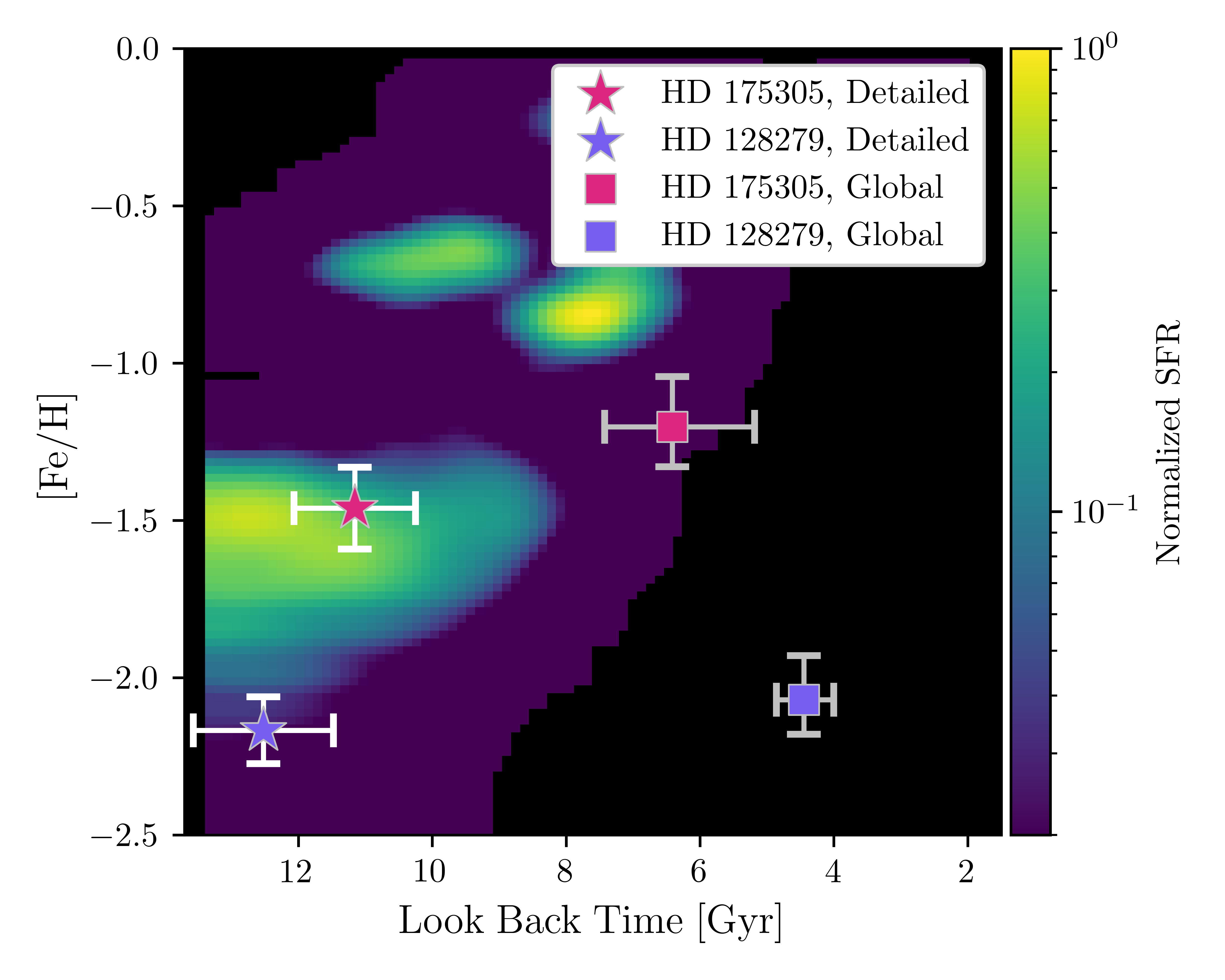}
    \caption{Normalized star formation rate in the age-[Fe/H] plane taken from the color magnitude diagram analysis of the HelmiA* sample in \citet{Ruiz-Lara_helmi_history} plotted on a log scale. The star symbols show the detailed asteroseismic modeling best fit age and [Fe/H] results for HD 175305 and HD 128279 while the square symbols show the results computed using only global asteroseismic parameters (\autoref{table:Grid_vs_Optimize_Results}).  }
    \label{fig:StarFormationHistoryHeatmapRuizLara_with_global}
\end{figure}

\autoref{fig:StarFormationHistoryFigRuizLara} shows that the best fit ages we derive for HD 175305 and HD 128279 (vertical lines in the top panel of \autoref{fig:StarFormationHistoryFigRuizLara}) match a period of high star formation rate calculated based on stringent membership cuts of the Helmi streams from \citet{Ruiz-Lara_helmi_history}. These ages, when combined with our best fit [Fe/H] relative to the cumulative metallicity distribution (see the bottom panel of \autoref{fig:StarFormationHistoryFigRuizLara}), indicate that both stars likely formed during the early to intermediate phases of their host dwarf galaxy's evolution, and prior to their accretion into the Milky Way. This result in summarized in \autoref{fig:StarFormationHistoryHeatmapRuizLara_with_global}, whereby our best fit parameters for HD 175305 lie directly within the region of strongest star formation [Fe/H] = -1.46 and age = 11.16 Gyr. Meanwhile, the older and more metal poor HD 128279 ([Fe/H] = -2.17 and age = 12.52 Gyr) lies in the old, low metallicity tail of star formation in the age-[Fe/H] plane. 

\subsection{Prospects for Characterizing the Helmi Streams' Progenitor}

With precise asteroseismic ages for more stream members in future work, we may begin to interpret the evolution of chemical trends across stream members. Previous spectroscopic studies of Helmi streams stars have pointed out that the stream members show a wide range in metallicity but lower $\alpha$-enhancement, as well as a tighter star-to-star range in $\alpha$-enhancement compared with the rest of the Milky Way halo \citep{Roederer2010, Gull2021, Matsuno2022, Horta2023}. These trends are indicative of complex star formation histories in the progenitor dwarf galaxy \citep{Matsuno2022}. In particular, \citet{Horta2023} found evidence that the [Mg/Fe] versus [Fe/H] gradient experiences a transition at a pivotal value of [Fe/H] $\sim 1.5$ dex at which the slope becomes shallower at higher metallicities. HD 175305 and HD 128279 both populate regions close to this transition. 

The different [$\alpha$/Fe] trends between the Helmi streams and halo stars are thought to originate from differences in star-formation efficiencies between both progenitor populations \citep{Matsuno2022}. \citet{Horta2023} found that the evolution of the magnesium (an $\alpha$-element) abundance as a function of [Fe/H] also indicates a burst of star formation, as their best fit for their sample of Helmi streams stars indicate an ‘inverted knee’, where the slope of the magnesium to iron abundance relationship becomes less negative at higher metallicities. Our results for HD 175305 and HD 128279 show that the older and lower metallicity star (HD 128279) is more enriched in $\alpha$-elements, including magnesium \citep{Ishigaki_2012}. With these new precise asteroseismic ages, it is possible to begin to measure the [$\alpha$/Fe] evolution over the history of the Helmi streams' progenitor, both in [Fe/H] and time. With only two stars with detailed asteroseismic ages, any detailed [$\alpha$/Fe]-age relationship results are still uncertain. Continued asteroseismic measurements from TESS \citep{TESS_inst} and future observations from PLATO \citep{Plato_mission, Plato2025} will be vital for more detailed characterization of the chemical evolution of the progenitor of the Helmi Streams. Taking the catalog of stars associated with the Helmi Streams from \citet{Dodd2023}, we approximate that $\sim50$ giant stars (absolute Gaia $G < - 2.8$) in the Helmi Streams will be sufficiently bright enough ($G<13$) for measuring at least the global asteroseismic parameters at a sufficiently high signal-to-noise. 

The Helmi streams have also been long studied for the presence of heavy element enrichment within its members (e.g., \citealt{Roederer2010, Limberg2021}). These are typically $r$-process elements like Europium, whose enrichment within stream members provide important clues to the presence of cataclysmic events like neutron star mergers or faster-acting magnetorotational supernovae within the progenitor dwarf galaxy. Determining the enrichment pathway between these two scenarios is necessary for reconstructing the star formation history of small galaxies, understanding their chemical evolution, determining neutron star merger delay times, and studying heavy element production on a wider scale. Analyzing Milky Way halo stars associated with two disrupted dwarf galaxies, \citet{Naidu_Ji_2022} found through a simple model analysis that $r$-process enhancement through neutron star mergers must be delayed by 0.5-1 Gyr. Asteroseismic ages of stars across different components of the halo with differing amounts of $r$-process elements will help constrain their origins by providing precise measurements of when such elements first became present across different sources. 

Based on measurements of europium and barium from \citet{Roederer2010}, HD 175305 is moderately $r$-process enhanced and categorized as type $r$-I (\citet{Frebel2018}) while HD 128279 is not $r$-process enhanced. This may suggest either different birth sites between the two stars within the progenitor, or potentially a $r$-process enrichment event between the birth of the two stars. The latter is difficult to conclude, given the similarity of the stars' ages, in consideration of their uncertainties. 

The abundances of $r$-process elements such as Thorium are used to determine cosmo-chronometric ages of $r$-process enhanced stars \citep{Butcher1987, Cayrel2001, Frebel2009} but when applying these techniques to HD 175305, \citet{Gull2021} derived an age of 0.5 Gyr, much younger than expected for a low mass giant star in a stellar stream. We note that the determination of precise asteroseismic ages for $r$-process enhanced stars like HD 175305 may provide a novel opportunities to calibrate such ages that are based on radioactive decay.



\subsection{Implications for the Use of Global Asteroseismology}

The masses and radii of tens of thousands of red giant stars have been determined using the asteroseismic scaling relations for $\nu_{\text{max}}$ \citep[$\nu_{\text{max}} \propto g \text{T}_{\text{eff}}^{-1/2}$,][]{Kjeldsen1995} and $\Delta \nu$ \citep[$\Delta \nu \propto \sqrt{\rho}$,][]{Brown1991}. \citet{LiTanda2022} found systematic offsets between the inferred masses and radii determined using the global asteroseismic parameters and the model solutions established using radial ($\ell = 0$) mode asteroseismology. Currently, only a few stars in the low-metallicity regime have been studied in detail using non-radial mode frequencies \citep{Chaplin2020, Huber2024, Larsen2025}. Using these scaling relations assumes that the structures of giant stars are homologous with respect to the Sun, which is not strictly true, especially as the scaling relations do not account for metallicity variances and evolutionary state. Additional verification is needed to assess the accuracy and reliability of scaling-relation derived masses and radii for low-metallicity evolved stars. 

Recent theoretical work has investigated how the scaling relation for $\nu_{\text{max}}$ varies with metallicity  using 3D simulations \citep{Zhou2024} but did not find that $\nu_{\text{max}}$ correlates with metallicity, in contrast with recent 1D stellar modeling work \citep{Huber2024, LiYaguang2024_realistic, Larsen2025}. These modeling studies indicate that the asteroseismic scaling relations generally overestimate the stellar mass, thereby underestimating the stellar age when compared with individual frequency modeling. We confirm these findings by directly applying the asteroseismic scaling relations for stellar mass, 
\begin{equation}
     \bigg( \frac{M}{\text{M}_\odot}\bigg) \approx \bigg(\frac{\nu_{\textrm{max}}}{\nu_{\textrm{max} \odot}} \bigg)^3  \bigg(\frac{\Delta \nu}{\Delta \nu_{\odot}} \bigg)^{-4} \bigg( \frac{T_{\textrm{eff}}}{\text{T}_{\textrm{eff} \odot}} \bigg)^{\frac{3}{2}} 
\end{equation}
sampling the observed $\nu_{\text{max}}$, $\Delta \nu$, and $T_{\text{eff}}$ within their corresponding uncertainties and find that the scaling relation masses for HD 175305 and HD 128279 are $1.03^{+0.07}_{-0.06} \text{M}_{\odot}$ and $1.02^{+0.06}_{-0.06} \text{M}_{\odot}$ respectively. Likewise, we use the scaling relation for stellar radius, 
\begin{equation}
     \bigg( \frac{R}{\text{R}_\odot}\bigg) \approx \bigg(\frac{\nu_{\textrm{max}}}{\nu_{\textrm{max} \odot}} \bigg)  \bigg(\frac{\Delta \nu}{\Delta \nu_{\odot}} \bigg)^{-2} \bigg( \frac{T_{\textrm{eff}}}{\text{T}_{\textrm{eff} \odot}} \bigg)^{\frac{1}{2}} 
\end{equation}
and find the scaling relation radii for HD 175305 and HD 128279 are $8.10^{+0.17}_{-0.17} \text{R}_{\odot}$ and $4.16^{+0.08}_{-0.08} \text{R}_{\odot}$ respectively. The scaling relation derived masses and radii for both HD 175305 and HD 128279 are larger than the associated best-fit model masses and radii from \autoref{table:Results}. 

Since the global asteroseismic parameters were not used in our individual frequency modeling, following \citet{ASFGRID1, LiYaguang2024_realistic, Huber2024, Larsen2025}, we can quantify the best fit model deviation from the $\nu_{\text{max}}$ scaling relation ($\nu_{\text{max}} \propto g / \sqrt{\text{T}_{\text{eff}}}$) using 
\begin{equation}
    f_{\nu_{\text{max}}} = \frac{\nu_{\text{max}}}{\nu_{\text{max}, \odot}} \bigg{(}\frac{M}{\text{M}_{\odot}}\bigg{)}^{-1} \bigg{(}\frac{R}{\text{R}_{\odot}}\bigg{)}^{2} \bigg{(}\frac{T_{\text{eff}}}{\text{T}_{\text{eff},\odot}}\bigg{)}^{1/2}  .
\end{equation}\label{eq:f_nu_max}
Using the measured $\nu_{\text{max}}$ values, as well as our best fit mass, radius, and temperature results from our detailed asteroseismic modeling of HD 175305 and HD 128279, we find that $f_{\nu_{\text{max}}} = 1.05$ for HD 175305 and $f_{\nu_{\text{max}}} = 1.11$ for HD 128279. These results are comparable with recent detailed asteroseismic studies of metal poor stars such as KIC 8144907 ($f_{\nu_{\text{max}}} = 1.056$), KIC 4671239, ($f_{\nu_{\text{max}}} = 1.101$), and KIC 7683833 ($f_{\nu_{\text{max}}} = 1.098$) \citep{Huber2024, Larsen2025}.



\begin{table*}[ht]
    \caption{Detailed asteroseismic modeling results for HD 128279 and HD 175305 compared with the results using global asteroseismic grid based modeling. }\label{table:Grid_vs_Optimize_Results}
    \centering
    \begin{tabular}{ccccccc}
        \toprule
        & \multicolumn{3}{c}{\textbf{HD 175305}} & \multicolumn{3}{c}{\textbf{\textbf{HD 128279}}} \\
        \midrule
        & Detailed & Global (Custom Grid) & Global (Asfgrid) & Detailed & Global (Custom Grid) & Global (Asfgrid)\\
        \midrule
        Mass ($\text{M}_{\odot}$)  &$0.83 \pm 0.02$ & $0.98^{+0.03}_{-0.03}$ & $0.92^{+0.03}_{-0.03}$  &$0.77 \pm 0.02$ & $1.08^{+0.03}_{-0.02}$ & $1.06^{+0.08}_{-0.05}$\\       
        Radius ($\text{R}_{\odot}$) &$7.40 \pm 0.07$ & $7.93^{+0.80}_{-0.12}$  &$7.69^{+0.06}_{-0.06}$  &$3.80 \pm 0.03$ & $4.26^{+0.06}_{-0.03}$ & $4.27^{+0.09}_{-0.11}$\\       
        Age (Gyr) & $11.16 \pm 0.91$ & $6.41^{+1.23}_{-1.01}$  & $8.73^{+1.32}_{-1.07}$  &$12.52 \pm 1.05$ & $4.45^{+0.45}_{-0.41}$ & $4.42^{+1.22}_{-0.67}$\\
        \bottomrule
    \end{tabular}
\end{table*} 

In order to further investigate how our detailed asteroseismic modeling age results for HD 175305 and HD 128279 compare with global asteroseismic results, we use a custom grid of stellar models calculated using the same model physics as the optimization tracks as well as Asfgrid \citep{ASFGRID1, ASFGRID2} to find masses and radii for our target stars based on the observed spectroscopic and global asteroseismic parameters. The custom grid consisted of 1024 model tracks with varying Mass ($0.7 \leq M \leq 1.2$), initial helium abundance ($0.23 \leq Y_0 \leq 0.29$), initial iron abundance ($-3 \leq \textrm{{[Fe/H]}} \leq 0.2$), and mixing length ($1.6\leq \alpha_{\text{mlt}} \leq 2.0$), with the initial parameters for each track selected using Sobol sampling \citep[as in][]{Lindsay2024}. The models for this grid were enhanced in alpha elements and were calculated using alpha element enhanced opacity tables ([$\alpha$/Fe] = 0.2) so no correction to the observed [Fe/H] was required \citep{salaris}.

\begin{figure*}
    \centering
    \includegraphics[width=0.8\textwidth]{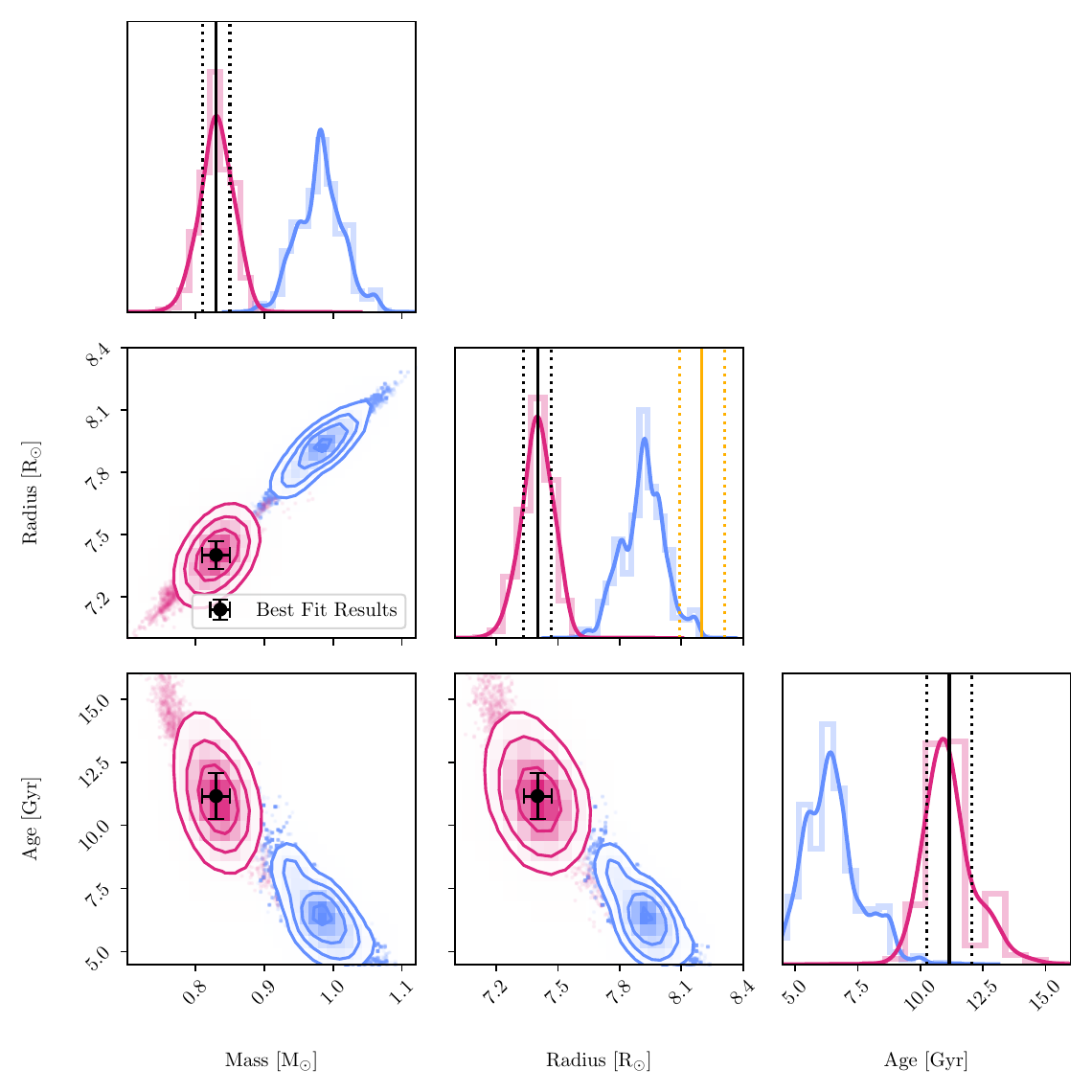}
    \caption{Global asteroseismic results for the stellar mass, radius, and age of HD 175305 determined with our custom grid, compared with the detailed asteroseismic optimization results. The blue points, histograms, and contours show the distribution of grid samples while the dark red points, histograms, and contours show the distribution of masses, radii, and ages for the best-fit model from each of the evolutionary tracks calculated as part of the optimization procedure. The black points and vertical lines in each plot show the results from the asteroseismic optimization procedure (\autoref{table:Results}). The vertical gold lines in the radius histogram show the interferometric radius for HD 175305 ($R = 8.2 \pm 0.11 \text{R}_{\odot}$) reported in \citet{Karovicova2020}.}
    \label{fig:Corner_HD175305}
\end{figure*}

\begin{figure*}
    \centering
    \includegraphics[width=0.8\textwidth]{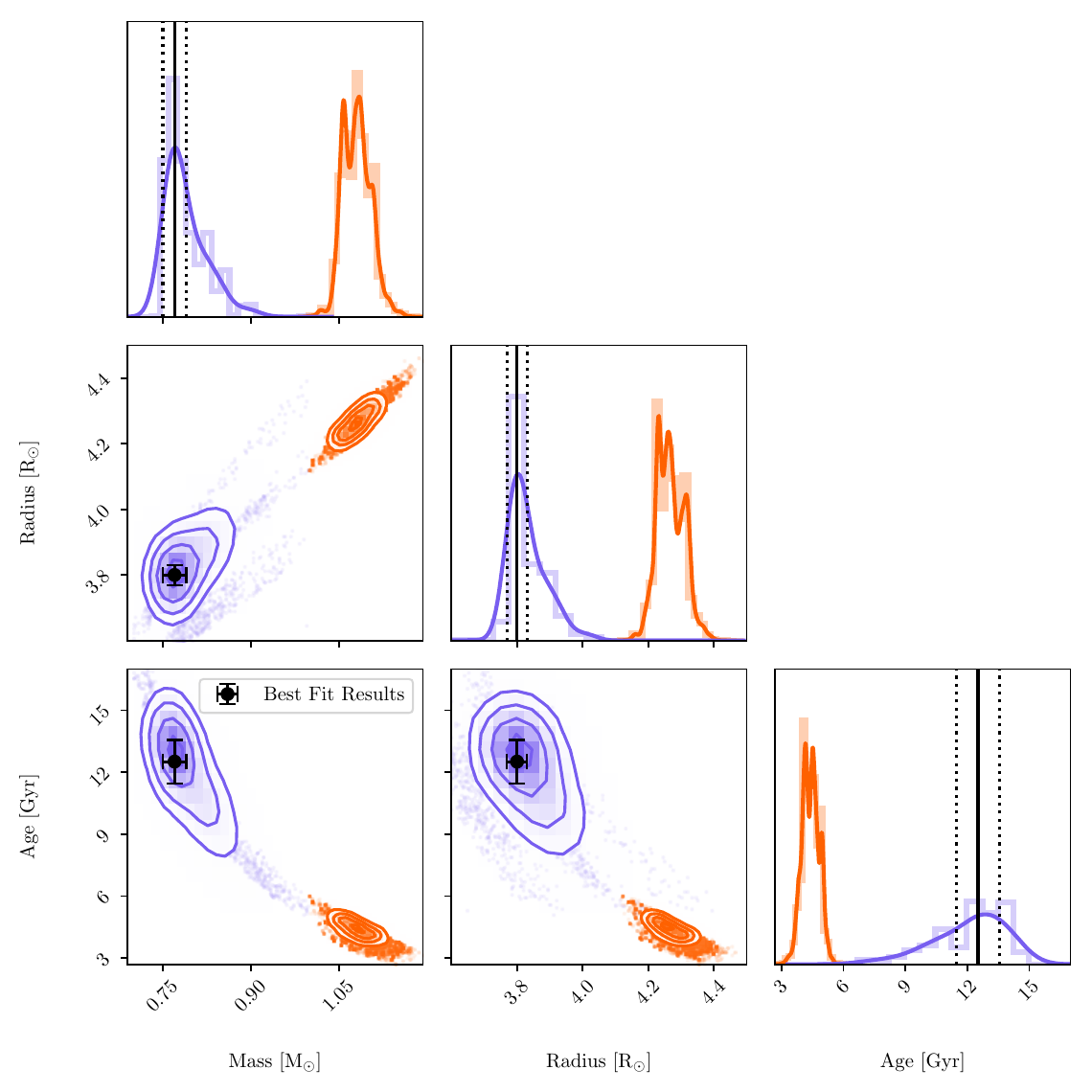}
    \caption{Global asteroseismic results for the stellar mass, radius, and age of HD 128279 compared with the detailed asteroseismic optimization results. The orange points, histograms, and contours show the distribution of grid samples while the purple points, histograms, and contours show the distribution of masses, radii, and ages for the best-fit model from each of the evolutionary tracks calculated as part of the optimization procedure. The black points and vertical lines in each plot show the results from the asteroseismic optimization procedure (\autoref{table:Results}). }
    \label{fig:Corner_HD128279}
\end{figure*}

For each star, the observed T$_{\text{eff}}$, [Fe/H], $\nu_{\text{max}}$, and $\Delta \nu$ (see \autoref{table:stellar_parameters}) are used as inputs to \texttt{modelflows} \citep{Hon_2024_modelflows}, which interpolates the grids of models by emulation using a neural network. The mass, radius, and age custom grid-based results for HD 175305 and HD 128279 are shown with corner plots in \autoref{fig:Corner_HD175305} and \autoref{fig:Corner_HD128279} and listed along with the best-fit results in \autoref{table:Grid_vs_Optimize_Results}. The grid-based parameter errors in \autoref{table:Grid_vs_Optimize_Results} are the 16th, 50th, and 84th mass, radius, and age percentiles from the grid samples. The best fit results from our detailed modeling are shown with black points and lines in each panel of  \autoref{fig:Corner_HD175305} and \autoref{fig:Corner_HD128279}. We note that the TACO-derived $\Delta \nu$ values for HD 175305 and HD 128279 were multiplied by 1.0109 before inputting them into \texttt{modelflows} to account for the difference between the observed $\Delta \nu$ and the frequency separation determined from non-surface-term-corrected models \citep{Viani_2018}. Since the stellar modeling codes used in this work and in \citet{Viani_2018} are not the same, we also tested the uncorrected $\Delta \nu$ value to confirm our results remain consistent with and without the $\Delta \nu$ correction.  

We find that the asteroseismic masses determined using the scaling relations are significantly higher than the best-fit masses we found using individual $\ell = 0, 1, \text{ and } 2$ mode frequencies (\autoref{table:Grid_vs_Optimize_Results}). We also find that this discrepancy is consistent whether the scaling relation results are derived from our custom grid of models or from the widely used Asfgrid models \citep{ASFGRID1, ASFGRID2}. The trend of scaling relation derived masses being larger than the best fit masses from detailed modeling agrees with previous detailed asteroseismic analyses of metal poor evolved stars \citep{Huber2024, Larsen2025} as well as previous work which found that stellar masses determined using the scaling relations were overestimated for metal-poor red giants \citep[e.g.][]{Epstein2014}. The agreement between the global asteroseismic mass and the mass obtained using individual mode frequencies is worse for HD 127289, which may indicate the mass discrepancy expands with lower metallicity and higher $\alpha$-enhancement. 

Mirroring the results for global asteroseismic stellar mass determinations, the lower right histogram panels of \autoref{fig:Corner_HD175305} and \autoref{fig:Corner_HD128279} show that the global asteroseismic stellar age results determined using grid based modeling are underestimated compared with the results from detailed asteroseismic modeling using individual mode frequencies. The best-fit model age for HD 175305 is about 4$\sigma$ larger compared with the age from global asteroseismology, while the best-fit model age for HD 128279 is much older (almost 18$\sigma$) compared with the age from global asteroseismology. The discrepancies between the masses and ages determined for HD 128279 are larger compared with the discrepancies between the masses and ages determined for HD 175305, potentially indicating that lower metallicity, more $\alpha$-element enhanced stars like HD 128279 suffer more from errors involved with leveraging the global asteroseismic scaling relations, however detailed asteroseismic analyses of more stars with varying metallicities and $\alpha$-abundances is needed to determine the scope and origin of the problem. 

\citet{apok2_1} present a catalog combining APOGEE spectra data and K2 asteroseismic data for 7500 evolved stars, including a large population of low-metallicity stars. As we find in this work, \citet{apok2_1} found that the asteroseismic masses determined for low metallicity stars using global asteroseismology were significantly larger than astrophysical estimates. \citet{apok2_1} argued this effect may be due to offsets in the adopted fundamental temperature scale for metal poor stars \citep{Gonzalez_Hernandez_2009} rather than metallicity-dependent issues with the asteroseismic scaling relations However, we note that we did not use the APOGEE effective temperatures in our detailed or global asteroseismic modeling and still find the masses determined using global asteroseismology are much larger than masses determined using detailed asteroseismology. 

The high mass, young age global asteroseismic results are also discrepant with color-magnitude diagram analysis. We compare the detailed and global asteroseismic results to the color-magnitude diagram fitting analysis of \citet{Ruiz-Lara_helmi_history} in \autoref{fig:StarFormationHistoryHeatmapRuizLara_with_global}. The younger grid based ages shown with square symbols in \autoref{fig:StarFormationHistoryHeatmapRuizLara_with_global} are less accordant with the areas of high star formation. Additionally, the older ages we derive using detailed asteroseismology for the Helmi streams members are consistent with the idea that star formation was quenched once the progenitor dwarf galaxy merged with the Milky Way approximately 5-8 Gyr ago \citep{Koppelman2018, Ruiz-Lara_helmi_history, Naidu2022} while the young age we obtain using grid based modeling without individual mode frequencies for HD 128279 is inconsistent with this quenching theory. 


We also find that the radii determined using global asteroseismic analysis are significantly larger than the radii of the best fit models we determined for HD 175305 and HD 128279. Using the SEDEX pipeline \citep{Yu_2023}, we found that spectral energy distribution based radii results for both HD 175305 ($R_{\textrm{SED}} = 7.6 \pm 0.7 \text{R}_{\odot}$) and HD 128279 ($R_{\textrm{SED}} = 3.9 \pm 0.4 \text{R}_{\odot}$) are consistent to 1$\sigma$ with both the detailed and global asteroseismic radii listed in \autoref{table:Grid_vs_Optimize_Results}. On the other hand, CHARA array interferometry of HD 175305 \citep{Karovicova2020} measured the star's radius to be $8.2 \pm 0.11 \text{R}_{\odot}$, significantly larger than the radii determined using either detailed or global asteroseismology (\autoref{fig:Corner_HD175305}). The interferometric radius of HD 175305 is also larger than the radius determined using SED fitting. 

Previous studies have found that stellar radii calculated using the global asteroseismic scaling relations agree with radii determined using interferometry to within $\sim5\%$ \citep[e.g.][]{Huber2012}. Recent detailed modeling work also found that asteroseismic radius of HD 219134 is significantly smaller than the interferometric radius \citep{YaguangLi2025}. Our findings are similar to those of \citet{YaguangLi2025} as we determine a best-fit model whose oscillation modes match the observed modes but whose radius is significantly smaller than radii determined using interferometry. \citet{Ash2025} found that different methods of determining correction factors used in the $\Delta \nu$ scaling relation cause differences in the asteroseismic scaling relation derived radii for luminous red giants. Thus, different implementations of a $\Delta \nu$ scaling relation correction may explain the discrepancy between the scaling relation derived radius we derive for HD 175305 and the interferometric radius. 

The discrepancy between the best-fit model radius and the interferometric radius for HD 175305 is larger than the discrepancy between the scaling relation derived radius and the interferometric radius. This may be due in part to the fact the surface term is fitted as a free parameter in our modeling, and we are therefore losing sensitivity to the stellar surface in our seismic constraint. Matching surface corrected model mode frequencies to observed oscillations does well at modeling the deep internal structure, but not the near-surface layers. Thus, our detailed modeling does not give the stellar radii as directly constrained by asteroseismology, but rather the stellar model radii whose interior structures agree with the asteroseismology. Improved agreement between asteroseismic best-fit model radii determined using individual mode frequencies and interferometric radii will rely on better models for the atmosphere of stars and new treatments for correcting stellar model oscillation frequencies for near-surface effects. We also note that the effective temperature of HD 175305 determined using interferometry \citep[$T_{\text{eff}} = 4850 \pm 118$ K][]{Karovicova2020} was significantly lower than the temperature we employed in our modeling. In this work we used the effective temperature from \citet{Ishigaki_2012} in order to stay consistent between our modeling for both HD 175305, where interferometric data is available, and HD 128279, where interferometric data is not available. 


\subsection{Potential Constraints from Gravity Mode Period Spacings}
\label{subsec:175305_State}
Thus far, we have assumed that HD 175305 and HD 128279 are stars on their first ascent up the red giant branch, such that they are burning hydrogen in a shell around an inert helium core. This assumption is made based on the two stars' kinematic membership in the relatively old Helmi streams structure and their low metallicities. In this section, we examine the assumption that HD 175305 and HD 128279 are first ascent red giant branch stars by examining the stars' gravity mode (g-mode) period spacing ($\Delta\Pi_{\ell}$) and other observable parameters. HD 128279 ($L \simeq 11 \text{L}_{\odot}$) has a much lower observed luminosity than HD 175305 ($L \simeq 33$), which definitely rules it out as a core-helium burning star, since standard helium burning red giants are much more luminous \citep[$L \gtrsim 40$,][]{Girardi_2001}. This is confirmed by the $\Delta \nu$ of HD 128279, which is too large for HD 128279 to be in any other evolutionary state besides the red giant branch.

$\Delta\Pi_{\ell}$ refers to the asymptotic period spacing between consecutive g-modes of the same spherical degree, $\ell$, and probes the core region of giant stars as $\Delta\Pi_{\ell}$ depends on the buoyancy frequency profile in the core, 
\begin{equation}
    \Delta \Pi_{\ell} = \frac{2 \pi^2}{\sqrt{\ell (\ell + 1}} \left( \int_{\textrm{core}} \frac{N}{r}dr\right)^{-1}
\end{equation}
where $N$ is the Brunt–Väisälä, $r$ is the radius coordinate, $\ell$ is the angular degree, and the integral is taken over the extent of the radiative core of the red giant \citep{Tassoul1980}. Measurements of $\Delta\Pi_{\ell = 1}$ are especially valuable for determining the evolutionary state and ages of red giants. Larger values of $\Delta\Pi_{\ell = 1}$ indicate a less compact core and are generally representative of core-Helium burning giants, while hydrogen-shell burning giants tend to have more compact, degenerate cores, leading to smaller $\Delta\Pi_{\ell = 1}$ values \citep{Bedding2011}. The boundary between these two evolutionary phases is clearly defined for low-mass ($M < 1.2 \text{M}_{\odot}$) red giant stars, whereby those with $\Delta\Pi_{\ell = 1} < 100\,$s are hydrogen-shell burning and those higher $\Delta\Pi_{\ell = 1}$ are core-helium burning stars \citep{Mosser2015, Vrard2016, Kuszlewicz_2023}. Beyond a determination of evolutionary phase, $\Delta\Pi_{\ell = 1}$ may also provide an additional asteroseismic constraint to include in stellar modeling. 

\begin{figure}
    \centering
    \includegraphics[width=0.47\textwidth]{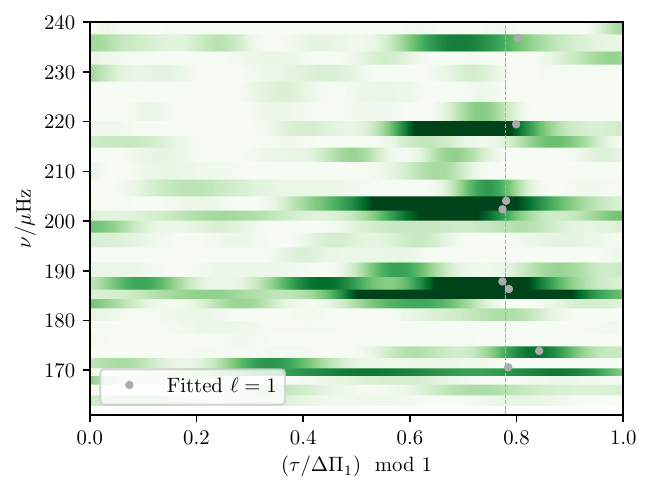}
    \caption{The \reggae-generated stretched period échelle power diagram (frequency versus the stretched period folded by the asymptotic period spacing) for HD 128279. Before plotting, the \pbjam-generated fit to the $\ell=0$ and $\ell=2$ modes divided out from the power spectra, so that only residual power from dipole modes is depicted. The gray points show the dipolar mode frequencies from \autoref{table:HD_128279_freqs} and the vertical gray dashed line shows the value of $\epsilon_\text{g}$ (0.78). }
    \label{fig:period_echelle}
\end{figure}
\begin{figure}
    \centering
    \includegraphics[width=0.47\textwidth]{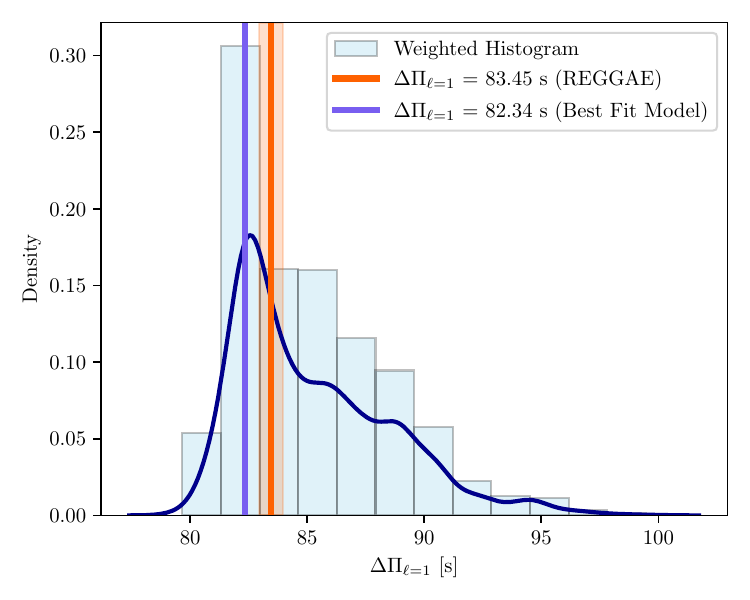}
    \caption{The likelihood weighted histogram of $\Delta \Pi_{\ell = 1}$ values from each of the best-fit models from every evolutionary track, calculated as part of the optimization procedure for HD 128279 (\autoref{appendix1}). The solid dark blue curve shows the kernel density estimate of the same likelihood weighted $\Delta \Pi_{\ell = 1}$ values. The overall best fit model $\Delta \Pi_{\ell = 1}$ value is shown with a vertical purple line, and the observed $\Delta \Pi_{\ell = 1}$ value obtained using \pbjam/\reggae is shown with the vertical orange line.}
    \label{fig:period_spacing}
\end{figure}

We were able to determine a period spacing for HD 128279, finding $\Delta\Pi_{\ell = 1} = 83.45 \pm 0.5$ seconds using the peakbagging tools \pbjam/\reggae \citep{PBJam, REGGAE}. This is done by first fitting a model for the $\ell = 0$ and $\ell = 2$ modes only against the power spectrum (using \pbjam version 2.0), which was divided out to produce a residual power spectrum containing power from only dipole-mode oscillations. \autoref{fig:period_spacing} shows this residual power spectrum, in the form of a stretched period-échelle power diagram (as constructed following \citealt{ong_mode_2023}), where the power spectrum is phase-folded by the g-mode period spacing $\Delta\Pi_1$, after applying a nonlinear coordinate transformation which undoes the coupling between the p- and g-mode cavities (e.g. \citealt{Mosser2015}). The asymptotic period spacing is determined using \reggae, taking a maximum likelihood estimate given a prior on the gravity mode phase shift ($\epsilon_\text{g}$) as being 0.78 (which is the value that emerges from asymptotic analysis of high-order dipole g-modes: \citealt{provost_asymptotic_1986}). \autoref{fig:period_echelle} shows that phase-folding by this value of $\Delta \Pi_{\ell = 1}$ results in the g-mode peaks forming a single vertically-aligned ridge along the vertical gray dashed line in the \'echelle power diagram. 

Given that HD 128278 has only been observed for two, non-contiguous sectors, the data is insufficient to simultaneously and independently constrain the pure p-mode frequencies (one parameter for each radial order), $\epsilon_{\text{g}}$, the coupling factor ($q$), and $\Delta \Pi_{\ell = 1}$. This results in a relatively high uncertainty for the observed $\Delta \Pi_{\ell = 1}$ value (0.5 seconds)  --- which is orders of magnitude less precise than those measured from Kepler, or from the TESS continuous viewing zones --- as the $\Delta \Pi_{\ell = 1}$ value is statistically degenerate with other g-mode parameters such as $\epsilon_{\text{g}}$. \autoref{fig:period_spacing} shows this empirical value of $\Delta \Pi_{\ell = 1}$ using the orange vertical line, compared to a histogram showing the $\Delta \Pi_{\ell = 1}$ values from each of the best-fit models along each of the tracks calculated as part of the asteroseismic optimization procedure. The best fit model we constructed for HD 128279 during the optimization procedure has an asymptotic period spacing value of $\Delta \Pi_{\ell = 1} = 82.34$ s (purple vertical line in the right panel of \autoref{fig:period_spacing}). Even though $\Delta \Pi_{\ell = 1}$ values were not explicitly included in the asteroseismic optimization procedure, the best fit models of HD 128279 roughly agree with the observed $\Delta \Pi_{\ell = 1}$ value. Including a $\Delta \Pi_{\ell = 1}$ constraint does not change the asteroseismic modeling results significantly, owing to the lack of precision in this constraint. 

We attempted to also use the peakbagging tools \pbjam/\reggae\ to derive $\Delta\Pi_{\ell = 1}$ from the power spectra of HD 175305, but were unable to do so with a high degree of confidence assuming HD 175305 is a first-ascent giant. This is due to the unavailability of a suitable Bayesian prior from \pbjam/\reggae due to a combination of the low-metallicity and advanced evolutionary state of HD 175305. We attempted to circumvent this issue by setting up many samplers to exhaustively sample the posterior distribution even in areas of low probability in an attempt to find a global maximum. Performing this 'umbrella' sampling using \pbjam/\reggae\ and the EMUS formulation of \cite{Matthews2018}, we obtain a highly multi-peaked posterior distribution for $\Delta\Pi_{\ell = 1}$ with an overall global maximum likelihood around $\Delta\Pi_{\ell = 1} = 275$ seconds. This large value of $\Delta\Pi_{\ell = 1}$ is not consistent with HD 175305 being a red giant branch star, though we note that there is a local maximum in the \pbjam/\reggae\ determined posterior around 68 seconds, which would be consistent with HD 175305 being a red giant branch star.  

Another method to measure the period spacing of a red giant is to remove the radial and quadrupole modes from a star's power spectrum to obtain just the mixed mode power spectrum. The mixed mode power spectrum is generally calculated by dividing the full power spectrum by a model power spectrum which only contains the $\ell = 0$ and $\ell = 2$ mode fits determined during the peakbagging process. Then the mixed-mode period spacing is found by locating the maximum of the Fourier transform of the stretched mixed-mode oscillation power spectrum, as described in \citet{Vrard2016}. Using this method, allowing trial $\Delta\Pi_{\ell = 1}$ values between 20 and 400 seconds and trial coupling factor ($q$) values between 0.0 and 0.4, we find a mixed-mode period spacing of  $\Delta\Pi_{\ell = 1} = $68.05 seconds and a coupling factor of $q = 0.11$. The estimated error on the period spacing (taking the uncertainty to be the time resolution of the Fourier spectrum of the stretched mixed-mode spectrum) is $\delta(\Delta\Pi_{\ell = 1})_{\text{res}} = \nu_{\text{max}}\Delta\Pi_{\ell = 1}^2 = 0.24$ seconds. 

In light of these conflicting period spacing results, we cannot definitively rule out HD 175305 being a core Helium burning star based solely on the asteroseismic data. If HD 175305 was a secondary clump helium burning star, it would be much younger, unless it reached its current structure through a mass transfer or merger event earlier in its evolutionary history \citep{Rui2021, Rui2024}. In this scenario, HD 175305, in its current state, would appear younger than its true age even when analyzed using asteroseismic techniques. The origins of young $\alpha$-rich stars have been studied in recent years through a combination of asteroseismic and spectroscopic techniques and binary interactions of older stars appear important to the existence of the seemingly young $\alpha$-rich stars \citep{Chiappini2015, Martin2015, Jofre2023, Yu2024}, however genuinely young $\alpha$-rich also appear \citep{Lu2025}. Young $\alpha$-rich stars are a minority, representing approximately 10\% of $\alpha$-enhanced field stars \citep{Grisoni2024} and in general, $\alpha$-rich helium burning stars are rare \citep{APOKASC3}. Therefore, there is a low chance HD 175305 is truly one of these objects, though additional asteroseismic measurements of the period spacing of HD 175305 would be able to determine the evolutionary state with more certainty.



\section{Summary and Conclusion}
We have conducted a detailed asteroseismic analysis of two halo star members of the Helmi streams \citep{Helmi1999}, which are remnants of a system believed to have merged with the Milky Way 5 to 8 billion years ago \citep{Kepley2007, Koppelman2018, Ruiz-Lara_helmi_history, Woudenberg2024}. These target stars were identified in \citet{Dodd2023} by analyzing the stellar abundance data with 6D Gaia position and velocity measurements and grouping stars to determine substructures in the local stellar halo. 

With data from TESS \citep{TESS_inst} we used the peak-bagging code TACO \citep{taco} to analyze the light curves of HD 175305 and HD 128279 and determined the global asteroseismic parameters ($\nu_{\text{max}}$ and $\Delta \nu$) as well as the individual radial, dipole, and quadrupole oscillation mode frequencies of the stars. The global asteroseismic parameters as well as the spectroscopic parameters for our target stars are listed in \autoref{table:stellar_parameters} and the mode frequencies for HD 175305 and HD 128279 are listed in \autoref{table:HD_175305_freqs} and \autoref{table:HD_128279_freqs} respectively. 

In order to determine the best fit parameters for these Helmi streams stars, we developed an asteroseismic modeling procedure which creates models using the stellar evolutionary code MESA, calculates the model oscillation frequencies using GYRE, and compares the spectroscopic and asteroseismic quantities of the models along an evolutionary track to the observed properties. The best fit model parameters for HD 175305 and HD 128279 are listed in \autoref{table:Results} and visualized in \autoref{fig:HD175305_results} and \autoref{fig:HD128279_results}, respectively. 

From our detailed asteroseismic modeling results, our takeaway points are:
\begin{enumerate}
    \item Grid based modeling using only global asteroseismic parameters is inadequate for accurately determining the stellar parameters of metal-poor stars in the Halo. Our best fit masses and radii for both HD 175305 ($M = 0.83 \text{M}_{\odot}$, $R = 7.40 \text{R}_{\odot}$) and HD 128279 ($M = 0.77 \text{M}_{\odot}$, $R = 3.80 \text{R}_{\odot}$) are lower than the masses and radii determine using the global asteroseismic parameters (\autoref{table:Grid_vs_Optimize_Results}). Additionally, the asteroseismic ages we determine for HD 175305 and HD 128279 are older compared with ages determined without individual mode frequencies, showing that detailed asteroseismology with individual modes is necessary to study metal-poor halo stars such as those associated with the Helmi streams, in agreement with previous asteroseismic studies \citep{Epstein2014, Huber2024, Larsen2025}. 
    \item The best fit asteroseismic ages and other parameters we determine for HD 175305 (11.16 Gyr) and HD 128279 (12.52 Gyr) are consistent with previous studies of the star formation histories of the Helmi streams \citep{Koppelman2019, Naidu2022, Ruiz-Lara_helmi_history}.
    \item  The old age of HD 128279 (12.52 Gyr) places a lower bound on when stars began to form in the progenitor of the Helmi streams.
    \item Future precise determinations of asteroseismic ages for more stars in Halo substructures including the Helmi streams may be combined with detailed chemical abundance measurements to more precisely determine the time scales at which $r$-process enrichment occurs in accreted dwarf galaxies.
\end{enumerate}
 
As we have shown, asteroseismology enables the precise dating of individual red giants in different components of the Milky Way's halo. As TESS continues to measure the flux of more halo stars and survey efforts produce precise chemical abundances for larger numbers of stars, additional detailed asteroseismic analyses of more halo stars will become possible, allowing for precise studies of the ages and stellar properties for many stars in different kinematic components of our galaxy. Future asteroseismic studies will produce precise ages for more stars in Milky Way substructures, especially with the advent of the Roman space telescope \citep{Roman_Asteroseismology} and the PLAnetary Transits and Oscillations of stars mission \citep[PLATO][]{Plato_mission}. Many precise ages of halo stars will therefore soon be available to calibrate the cosmochronometry of stellar streams and date the processes which enrich galaxies with heavy elements.

\section*{Acknowledgments}
CJL acknowledges support from a Gruber Science Fellowship and NSF grant AST-2205026. M.H. acknowledges support from NASA grant 80NSSC24K0228. JMJO acknowledges support from NASA through the NASA Hubble Fellowship grant HST-HF2-51517.001, awarded by the Space Telescope Science Institute (STScI). RAG and DBP acknowledge the support from the GOLF and PLATO Centre National D'{\'{E}}tudes Spatiales grants. TRL acknowledges support from Juan de la Cierva fellowship (IJC2020-043742-I) and Ram\'on y Cajal fellowship (RYC2023-043063-I, financed by MCIU/AEI/10.13039/501100011033 and by the FSE+). AH is grateful for financial support through a Spinoza Grant from the Dutch Research Council (NWO). We also thank Yaguang Li, Dan Huber, Sarbani Basu, Yasmeen Asali, William Cerny, the anonymous reviewer, as well as Earl Bellinger and the Yale AstroML group for their useful and constructive discussion. We acknowledge the use of TESS High-Level Science Products (HLSP) produced by the Quick-Look Pipeline (QLP) at the TESS Science Office at MIT, which are publicly available from the Mikulski Archive for Space Telescopes (MAST) at STScI. The specific observations analyzed can be accessed via \dataset[https://doi.org/10.17909/qps3-3v44]{https://doi.org/10.17909/qps3-3v44}. Funding for the TESS mission is provided by the NASA Explorer Program. STScI is operated by the Association of Universities for Research in Astronomy, Inc., under NASA contract NAS 5–26555. 

\software{
MESA \citep{Paxton2011,Paxton2013,Paxton2015,Paxton2018,Paxton2019,Jermyn2023}, GYRE \citep{Townsend2013}},
SciPy \citep{scipy}, Pandas \citep{pandas}, Astropy \citep{Astropy1, Astropy2, Astropy3}, Lightkurve \citep{Lightkurve}, \pbjam \citep{PBJam}, \reggae \citep{REGGAE}, TACO \citep{taco}, YABOX \citep{Yabox}, Asfgrid \citep{ASFGRID1, ASFGRID2}, \texttt{usample} \citep{Matthews2018}

The MESA and GYRE inlists we used to generate our models and calculate model frequencies are archived on Zenodo and can be downloaded at 
 \dataset[zenodo.org/records/15102877]{zenodo.org/records/15102877}. We also include the MESA evolutionary tracks and profiles for our best-fit models.


\bibliographystyle{aasjournalv7}
\bibliography{biblio}

\appendix 
\section{Cost Function Evaluation Steps}
\label{appendix1}

\begin{enumerate}
    \item \textbf{MESA Evolutionary Track:} For each iteration of the optimization, we first calculate a stellar model track using MESA version r22.05.1 \citep{Paxton2011,Paxton2013,Paxton2015,Paxton2018,Paxton2019,Jermyn2023} using a set of model parameters ($M_0$, $Y_0$, $f = Z_0 / X_0$, and $\alpha_{\text{mlt}}$) determined by the differential evolution algorithm. The initial mass, initial helium mass abundance, initial metal mass abundance divided by initial hydrogen mass abundance, and convective mixing length, are varied in the following ranges, $0.7 \leq M_0 \leq 1.0$, $0.245 \leq Y_0 \leq 0.27$, $0.0001 \leq f \leq 0.003$, and $1.6 \leq \alpha_{\text{mlt}} \leq 2.0$. The model element mixtures are enhanced in $\alpha$-elements from their GS98 values \citep{GS98} according to the values from \citet{Ishigaki_2012}, and are computed using correspondingly $\alpha$-element enhanced OPAL/Opacity Project opacity tables, included in MESA's kap module. Since the modeling in this work is restricted to low-mass hydrogen burning, we use MESA's default 'basic.net' nuclear network in our stellar evolutionary model calculations. MESA's equation of state module computes the thermodynamic properties of the stellar material by combining different input EOS sources including FreeEOS, OPAL/SCVH EOS, HELM, and SKYE EOS depending on the conditions of the stellar material \citep{Jermyn2023}.
    
    Elemental diffusion is incorporated \citep{Thoul1994}, along with a small amount of exponential convective overshoot underneath the convective envelope ($f_{\text{ov, exp}} = 0.01 $ and $f_0$ = 0.0005). Overshoot is applied because most current models of low mass giant stars do not reproduce the observed location of the red giant branch luminosity bump, and incorporating envelope overshoot helps to resolve this discrepancy \citep{Alongi1991, Khan2018, Lindsay2022}. All models use a varying Eddington gray atmospheric boundary condition, and the mixing length prescription of \citep{CoxGiuli1968}. 
    
    Along the stellar model's evolution, we save the stellar structure files along the part of the evolutionary track which coincides well with the target star's temperature and luminosity (within 10 $\sigma$ from the values in \autoref{table:stellar_parameters}). In order to perform asteroseismic modeling, mode frequencies must be calculated for all the models one would like to compare with the observations. In order to save time during the optimization process, we only calculate mode oscillation frequencies for these models which show good agreement with the values from \citet{Ishigaki_2012}. 
    
    \item \textbf{$\chi^2_{\text{spectroscopic}}$:} $\chi^2_{\text{spectroscopic}}$ is calculated by comparing the model effective temperature, luminosity, and [Fe/H] to the literature values for HD 175305 and HD 128279 listed in \autoref{table:stellar_parameters}. The overall $\chi^2_{\text{spectroscopic}}$ is calculated as:
    \begin{equation}
    \label{eq:chi2spec}
    \chi^2_{\text{spectroscopic}} = \chi^2_{\text{Luminosity}} + \chi^2_{\text{Temperature}} + \chi^2_{\text{[Fe/H]}}
    \end{equation}
    with each spectroscopic parameter, $P$'s, corresponding $\chi_{P}^2$ value calculated as:
    \begin{equation}
    \label{eq:chi2}
    \chi_{P}^2 = \frac{(P_{\text{obs}} - P_{\text{model}})^2}{\sigma_{P_{\text{obs}}}^2}.
    \end{equation}
    We measure [Fe/H] for each model along the evolutionary track by taking the outermost model cell metal and hydrogen abundances from MESA, then calculating the iron abundance compared with the solar value \citep{GS98}, taking into account the different relative abundances of iron in the metal mixture given the alpha-enhancement values of HD 175305 and HD 128279.
    
    \item \textbf{Calculate Model Mode Frequencies and $\chi^2_{\text{seismic}}$:} We then calculate the $\ell = $ 0, 1, and 2 oscillation frequencies for the models along the evolutionary track with low values of $\chi^2_{\text{spectroscopic}}$ using the stellar oscillation code GYRE \citep{Townsend2013}. We do this in steps, first calculating the $\ell = 0$ p-mode frequencies for all models with $\chi^2_{\text{spectroscopic}} < 100$. For models with $\ell = 0$ mode frequencies which are a good match to the observed $\ell$ = 0 modes, we calculate the $\ell = 2$ p-modes as well as the $\ell = 1$ mixed-modes following the mode isolation construction of \citet{ong2020}. 

    We match each model $\ell = 0$ of $\ell = 2$ p-mode frequencies to the observed frequencies based on their inferred values of radial order $n_\text{p}$. We infer the observed modes' $n_p$ values by plugging in the observed mode frequencies ($\nu_{\text{obs}}$) and the observed $\Delta \nu$ to the rearranged asymptotic expression for p-mode frequencies, $n_{\text{p}} = (\nu_{\text{obs}}/\Delta \nu) - (\ell /2) $. The model $n_\text{p}$ values are returned from GYRE. Since multiple $\ell = 1$ mixed-modes can have the same $n_\text{p}$ value, we match the observed and model $\ell = 1$ modes using a nearest-neighbor search. 
    
    Before calculating $\chi^2_{\text{seismic}}$, the model mode frequencies must be corrected for the `surface term,' which is a frequency-dependent frequency offset between models and stars caused by our inability to model the near-surface layers of a star properly. The $\ell = 0$ and 2 modes are corrected for the surface term using the two-term prescription from \citet{bg14}, with the $\ell = 0$ modes used to determine the coefficients of the two-term surface correction. The $\ell = 1$ mixed modes must be treated differently, as the g-mode component of a mixed mode does not suffer from a surface effect as these modes do not reach the surface like p-modes do. We compute the $\ell = 1$ pure-p- and g-mode frequencies, and associated coupling matrices using the $\pi$/$\gamma$ isolation method of \citet{ong2020}. The surface term correction is applied only to the $\pi$-mode component of the mixed mode, and the surface-term-corrected mixed mode frequencies are determined using the corrected $\pi$-mode frequencies \citep[see also the methods section of][]{Lindsay2024}.

    With the surface corrected mode frequencies, the seismic $\chi^2$ per degree of freedom, $\chi^2_{\text{seismic}}$, is calculated for models along each evolutionary track as:
    \begin{equation}
    \label{eq:chi2seis}
    \chi^2_{\text{seis}} =  \frac{1}{N_\nu - 1 - 2}\sum^{N_{\nu}}_{n} \frac{(\nu_{\text{obs}, n} - \nu_{\text{model}, n})^2}{\sigma^2_{\nu_{\text{obs}, n}} }.
    \end{equation}
    with the total number of modes $N_{\nu}$ and with each observed mode frequency, $\nu_{\text{obs}, n}$, and corresponding mode error $\sigma^2_{\nu_{\text{obs}, n}}$, matched to the corresponding surface corrected model mode frequency, $\nu_{\text{model}, n}$.

    \item \textbf{Account for limitations involved with correcting model frequencies using a power-law based frequency shift:} In the previous step, surface-term corrected frequencies are compared with the observed mode frequencies to find $\chi^2_{\text{seismic}}$, however we need to account for models that are not good fits but appear to match the observed frequencies due to surface term corrections. Near surface modeling errors cause higher frequency model modes to be larger than the corresponding mode frequencies, and this frequency difference increases at higher frequencies. To ensure the cost function prefers models with better fits we follow \citet{BasuKinnane2018}, \citet{Ong2021a}, \citet{Cunha2021}, and \citet{Lindsay2024} in accounting for this by adding another penalty function to $\chi^2_{\text{seis}}$ calculated from the 2 lowest frequency radial and quadrupole modes. 

    \begin{equation}
    \label{eq:low_n_cost}
    \chi^2_{\text{low }n} = \frac{1}{100} \sum_{\ell\in \{0,2\}} \frac{1}{2} \sum_{n = 0}^1\frac{(\nu_{\text{obs}, n} - \nu_{\text{uncorr model}, n})^2}{\sigma^2_{\nu_{\text{obs}, n}} }.
    \end{equation}

    
    
    \item \textbf{Return $\chi^2_{\text{total}}$:} The sum of each models' $\chi^2_{\text{spectroscopic}}$, $\chi^2_{\text{seismic}}$, and $\chi^2_{\text{low }n}$ is then taken to be $\chi^2_{\text{total}}$, and the best-fit model along the evolutionary track is taken to be the model with the lowest $\chi^2_{\text{total}}$. That lowest $\chi^2_{\text{total}}$ value is the output of the cost function which the differential evolution algorithm is minimizing, and used to choose subsequent combinations of input model parameters. After one cost function evaluation finishes, the process starts again with a new MESA evolutionary track. 
\end{enumerate}

\end{document}